\documentclass[onecolumn,superscriptaddress,nofootinbib]{revtex4-1}
\pdfoutput=1

\usepackage{graphicx,color}
\usepackage{amsfonts}
\usepackage{amssymb}
\usepackage{amsmath}
\usepackage{amsthm}

\newcommand{\HH}{H}

\let\oldmarginpar\marginpar
\renewcommand\marginpar[1]{\-\oldmarginpar[\raggedleft\tiny #1]%
{\color{red}\raggedright\tiny #1}}

\newcommand{\bra}[1]{\big<#1|}
\newcommand{\ket}[1]{|#1\big>}

\newcommand{\rem}[1]{}

\def\set#1{\{ #1 \}}
					
\begin{document}

\title{Statistical mechanics of classical and quantum computational complexity}

\author{C. R. Laumann}
\affiliation{Department of Physics, Princeton University, Princeton, NJ 08544,
  USA} 

\author{R. Moessner}
\affiliation{Max Planck Institut fur Physik Komplexer Systeme, %
	01187 Dresden, Germany}

\author{A. Scardicchio}
\affiliation{Abdus Salam International Centre for %
Theoretical Physics, Strada Costiera 11, 34014 Trieste, Italy}

\author{S. L. Sondhi}
\affiliation{Department of Physics, Princeton University, Princeton, NJ 08544}

\begin{abstract} 
The quest for quantum computers is motivated by their potential
for solving problems that defy existing, classical, computers. The theory of
computational complexity, one of the crown jewels of computer science, provides a
rigorous framework for classifying the hardness of problems according to the
computational resources, most notably time, needed to solve them. Its extension to
quantum computers allows the relative power of quantum computers to be analyzed.
This framework identifies families of problems which are likely hard for classical
computers (``NP-complete'') and those which are likely hard for quantum computers
(``QMA-complete'') by indirect methods. That is, they identify problems of
comparable worst-case difficulty without directly determining the individual
hardness of any given instance. Statistical mechanical methods can be used to
complement this classification by directly extracting information about particular
families of instances---typically those that involve optimization---by studying
random ensembles of them. These pose unusual and interesting (quantum)
statistical mechanical questions and the results shed light on the difficulty of
problems for large classes of algorithms as well as providing a window on the contrast
between typical and worst case complexity. In these lecture notes we present an
introduction to this set of ideas with older work on classical satisfiability and
recent work on quantum satisfiability as primary examples. We also touch on the
connection of computational hardness with the physical notion of glassiness. 
\end{abstract}

\maketitle

\section{Introduction} 
\label{sec:Intro}

A large and exciting effort is underway to build quantum computers. While the roots
of this effort lie in the deep insights of pioneers such as Feynman and
Deutsch, what triggered the growth was the discovery by Shor that a quantum computer
could solve the integer factoring problem efficiently --- a feat currently beyond
the reach of classical computers. In addition to the desire to create useful devices
intrinsically more powerful than existing classical computers, the challenge of
creating large quantum systems subject to precise control has emerged as the central
challenge of quantum physics in the last decade.

The first and primary objective - that of enhanced computational power - has in turn
spurred the founding of quantum computer science and the development of a rigorous
theory of the (potential) power of quantum computers: quantum complexity theory.
This theory builds on the elegant ideas of classical complexity theory to
classify computational problems according to the resources needed to solve them as
they become large. The distinction between  polynomial scaling of resources,
most notably time, and super-polynomial scaling (e.g. exponential) generates a
robust distinction between easy and hard problems.

While this distinction is easily made in principle, in practice complexity theory
often proceeds by the powerful technique of assigning guilt by association: more
precisely, that of classifying problems by mapping between them. This allows the
isolation of sets of problems that encapsulate the difficulty of an entire class:
for example, the so-called satisfiability problem (SAT) captures the difficulty of
all problems whose solution can be easily checked by a classical computation; the
quantum satisfiability problem (QSAT) does the same for quantum computers, as we
will explain later in these notes. The solution of these problems would thus enable
the solution of the vast set of of all checkable problems. This implication is a
powerful argument that both SAT and QSAT must be hard.

This kind of indirect reasoning is very different from the way physicists normally
approach problems: one of the purposes of these notes
is to help physics readers appreciate the power of the computer science approach.
However, the direct approach of examining actual problem instances and attempting to
come up with algorithms for them is, of course, also important and this is where
physicists are able to bring their own methods to bear. Specifically, physicists
have applied themselves to the task of trying to understand problems such as the two
satisfiability problems. These can be expressed as optimization problems and thus
look much like the Hamiltonian problems the field is used to. Even more
specifically, they have studied ensembles of these problems with a variety of
natural probability measures in order to reveal features of the hard instances.

As is familiar from the statistical mechanical theory of disordered systems such as
spin glasses, studying a random ensemble brings useful technical simplifications
that allow the structure of a typical instance to be elucidated with less trouble
than for a specific capriciously picked instance. This has enabled the
identification of phase transitions as parameters defining the ensembles are changed
- exactly the kind of challenge to warm a statistical physicist's heart. A further
major product of such work, thus far largely in the classical realm, has been the
identification of obstacles to the solution of such typical instances by large
classes of algorithms and the construction of novel algorithms that avoid these
pitfalls. We note that this focus on typical instances also usefully complements the
standard results of complexity theory which are necessarily controlled by the worst
cases - instances that would take the longest to solve but which may be very
unusual. This is then an independent motivation for studying such ensembles.
 
Certainly, the flow of ideas and technology from statistical mechanics to
complexity theory has proven useful. In return, it is useful to reflect on what
complexity theory has to say about physical systems. Here the central idea -
whose precise version is the Church-Turing hypothesis - is that a physical
process is also a computation in that it produces an output from an input. Thus
if a complexity theoretic analysis indicates that a problem is hard, any
physical process that encodes its solution must take a long time. More
precisely, the existence of hard optimization problems implies the existence of
a class of Hamiltonians whose ground states can only be reached in a time that
scales exponentially in the volume of the system {\it irrespective} of the
processes used. 

This sounds a lot like what physicists mean by glassiness. We remind the
reader that physical systems typically exhibit symmetric, high temperature or
large quantum fluctuation phases, with a characteristic equilibration time
that is independent of the size of the system. At critical points, or in
phases with broken continuous symmetries, algebraic dependences are the norm.
But glassy systems exhibit much slower relaxation and thus present a challenge
to experimental study in addition to theoretical understanding. Indeed, there
is no settled understanding of laboratory glassiness. Consequently, the
complexity theoretic arguments that imply the existence of glassy
Hamiltonians, both in the classical and quantum cases, ought to be interesting
to physicists. That said, we hasten to add that the connection is not so
simple for two reasons. First, complexity theoretic results do not always hold
if we restrict the degrees of freedom to live in Euclidean space - say on
regular lattices - and require spatial locality. Thus hard Hamiltonians in
complexity theory can look unphysical to physicists. Nonetheless, there are many
interesting low dimensional (even translation invariant) problems which are
hard in the complexity theoretic sense
\cite{Barahona:1982p157,Aharonov:2007fk}. Second, the physical processes
intrinsic to a given system can sometimes be slow for reasons of locality or
due to energetic constraints which are ignored when one is considering the full
set of algorithms that can solve a given optimization problem. Still, we feel
this is a direction in which Computer Science has something to say to Physics
and we refer readers to a much more ambitious manifesto along this axis by
Aaronson \cite{Aaronson:2005p9879} for stimulation.

In the bulk of these notes we provide an introduction to this complex
of ideas, which we hope will enable readers to delve further into the
literature to explore the themes that we have briefly outlined above. In the
first part, we provide a tutorial on the basics of complexity theory
including sketches of the proofs of the celebrated proofs of NP and QMA
completeness for the satisfiability and quantum satisfiability problems. In the
second part, we show how statistical methods can be applied to these problems and
what information has been gleaned. Unsurprisingly, the quantum part of this
relies on recent work and is less developed than the classical results. In the
concluding section we list some open questions stemming from the quantum
results to date. 


\section{Complexity theory for physicists} 
\label{sec:complexity_theory_for_physicists}

Complexity theory classifies how ``hard'' it is to compute the
solution to a problem as a function of the \emph{input size} $N$
of the problem instance. As already mentioned above, algorithms are considered
\emph{efficient} if the amount of time they take to complete
scales at most polynomially with the size of the input and
\emph{inefficient} otherwise. The classification of algorithms by
asymptotic efficiency up to polynomial transformations is the key
to the robustness of complexity theoretic results, which includes the
independence from an underlying model of computation.

In this line of reasoning, if P $\ne$ NP, as is the current
consensus, there are natural classes of problems which cannot be
solved in polynomial time by any computational process, including
any physical process which can be simulated by computer. Complexity theory
provides its own guide to focusing our attention on a certain set of NP
problems, those termed NP-complete, which capture the full hardness of the
class NP. In particular, the problem of Boolean satisfiability of 3-bit
clauses, 3-SAT, is NP-complete and therefore can encode the full
hardness of the class NP.

The advent of the quantum computer modifies the above reasoning
only slightly. It appears that quantum computers are somewhat
more powerful than their classical counterparts, so that we must
introduce new quantum complexity classes to characterize them.
Nonetheless, analogous statements hold within this new framework:
quantum polynomial (BQP) is larger than classical polynomial (P)
but not powerful enough to contain classical verifiable (NP), nor
quantum verifiable (QMA). If we wish to study the particularly hard 
quantum problems, we may turn to the study of QMA-complete
problems such as LOCAL HAMILTONIAN and the closely related QSAT. 

In this section, we provide a concise review of the key
concepts for the above argument culminating in a discussion of
worst-case hardness and the Cook-Levin theorem, showing the
existence of NP-complete problems, and the quantum analogues due
to Kitaev and Bravyi. This story motivates and complements the
statistical study of `typical' instances of 3-SAT, 3-QSAT and
other classical and quantum hard optimization problems, which
are discussed in the following sections. 

\subsection{Problems, instances, computers and algorithms} 
\label{sub:problems_instances_computers_and_algorithms}

The success of complexity theory as a classification scheme for the
``hardness'' of problems is in part due to the careful definitions employed.
Here we sketch the most important aspects of these concepts and leave the
rigorous formalism for the textbooks, of which we particularly recommend Arora
and Barak to the interested reader \cite{Arora:2009zv}. We have taken a
particular path through the forest of variations on the ideas below and do not
pretend to completeness.

Throughout these notes we focus on so-called \emph{decision
problems}, that is Yes/No questions such as ``Does the Hamiltonian $H$ have a
state with energy less than $E$?'' rather than more general questions like
``What is the ground state energy of $H$?'' This restriction is less dramatic
than it seems -- many general questions may be answered by answering a
(reasonably short) sequence of related Yes/No questions, much like playing the
game twenty questions -- and it significantly simplifies the conceptual
framework we need to introduce. Moreover, many of the essential complexity
theoretic results arise already within the context of decision problems.

A decision problem, then, is a question that one may ask of a class of
\emph{instances}. For example, the DIVIDES problem asks ``Does $a$ divide
$b$?'' for integers $a$ and $b$. Leaving the variables $a$ and $b$
unspecified, we clearly cannot yet answer this question. An \emph{instance} of
DIVIDES might be ``Does 5 divide 15?'' A moment's thought now reveals that a
definitive answer exists: Yes. We refer to instances of a problem as
Yes-instances (No-instances) if the answer is Yes (No). We follow
computer science convention by giving problems fully capitalized names.%
\footnote{We trust this will not give physics readers PROBLEMS.}

What does it mean to solve a problem? We would not feel we had solved the
problem if we could only answer a few specific instances. On the other hand,
we certainly could not expect to have a book containing the (infinite) table
of answers to all possible instances for easy reference. Thus, we want a
general \emph{algorithm} which, given an arbitrary instance, provides us with
a step-by-step recipe by which we can compute the answer to the instance. Some
physical object must carry out the algorithmic instructions and it is this
object that we call a \emph{computer} -- whether it is a laptop running C
code, ions resonating in an ion trap or a sibling doing long division with
pencil and paper. Thus, a solution to a decision problem is an algorithm which
can decide arbitrary instances of the problem when run on an appropriate
computer. Such an algorithm for a decision problem is often called a
\emph{decision procedure}.

Clearly it is less work to answer the DIVIDES problem for small numbers than
for large. ``Does 5 divide 15?'' takes essentially no thought at all while
``Does 1437 divide 53261346150?'' would take a few moments to check. We
therefore define the \emph{input size} (or just \emph{size}) of an instance as
the number of symbols we need to specify the instance. In the DIVIDES problem,
we could take the size as the number of symbols needed to specify the pair
$(a,b)$. The size $N$ of $(5,15)$ would be 6 while that of
$(1437,53261346150)$ is 18. 

Computer scientists measure the \emph{efficiency} of an algorithm by
considering the asymptotic scaling of its resource consumption with the input
size of the problem. More precisely, consider the finite but large collection
of all possible problem instances whose size is no greater than $N$. For each
of these instances, the algorithm will take some particular amount of time.
For the finite collection at size $N$, there will be a worst-case instance
which takes more time $T$ than any of the others at that size. Complexity
theory generally focusses on the scaling of this worst-case time $T$ as a
function of input size $N$ as $N \to \infty$. Clearly, the slower the growth
of $T$ with $N$, the more efficient the algorithm for large inputs. Indeed,
algorithmic procedures are considered efficient so long as $T$ grows at most
polynomially with $N$, for any particular polynomial we like. Thus both linear
and quadratic growth are efficient, even though linear growth clearly leeds to
faster computations, at least for sufficiently large $N$.
Anything slower, such as $T = O(e^N)$, is inefficient.

For example, the most famous decision procedure for DIVIDES -- long division
-- takes of order $T = O(\log b \times \log a) \le O(N^2)$ arithmetic steps to
perform the division and check the remainder is 0. That $T$ grows with $a$ and
$b$ logarithmically corresponds nicely to our intuition that bigger numbers
are harder to divide, but not too much harder. It is instructive to consider a
different, inefficient, algorithm for the same problem. Suppose we had not yet
learned how to divide but knew how to multiply. We might try the following
decision procedure: try to multiply $a$ with every number $c$ from $1$ up to
$b$ and check if $ac = b$. This trial-and-error approach would take $T =
O(\log a \times \log b \times b) \approx O(e^{cN})$ to try out all the
possibilities up to $b$. Even for relatively small instances, this approach
would quickly become prohibitively time consuming -- simply enumerating all of
the numbers of up to 30 digits at one per nanosecond would take longer than
the age of the universe!

Finally, we lift our classification of algorithm efficiency to a
classification of problem hardness: A problem is \emph{tractable} if there
exists an efficient algorithm for solving it and it is \emph{intractable}
otherwise. By this definition, DIVIDES is tractable (long division solves it
efficiently), despite the existence of alternative slower algorithms.
As we will discuss further in the following sections, we can rarely
\emph{prove} that no efficient algorithm exists for a given problem, but
complexity theory nonetheless offers strong arguments that certain large
classes of problems are intractable in this sense.

Computers are clearly central to the determination of the difficulty of
problems -- we classify problems according to the efficiency of the
computational algorithms that exist for treating them. In addition to the time
taken, we can measure the resource requirements in various implementation
dependent ways -- memory consumed, number of gates required, laser pulse
bandwidth, quantity of liquid Helium evaporated. One might
expect that the kind of computer that we use would greatly influence any
complexity classification. At the very least, your laptop will be faster than
your brother at dividing 1000 digit numbers. The beauty of the definition of
efficiency by polynomial scaling is that many of these implementation
dependent details drop out and we really can focus on the time
efficiency as an overall measure of difficulty.\footnote{In practice the
amount of memory or the number of cores in a workstation regularly limits its
ability to do computations. Since in finite time, even a parallel computer can
only do a finite amount of work or address a finite amount of memory, a
polynomial bound on $T$ also provides a polynomial bound on the space
requirements. Likewise, finite parallelization only provides constant time
improvements. More refined classifications can be made by restricting resource
consumption more tightly but we will not consider them here.}

The robustness of these definitions follows from one of the great ideas of
complexity theory: up to polynomial overheads, any reasonable classical
computer may be simulated by any other. This is known as the strong
Church-Turing hypothesis and, as its name suggests, is only a conjecture.
Nonetheless, it has been examined and confirmed for many particular models of
classical computation\footnote{For example, Turing machines, Boolean circuit
models and your laptop.} and is widely believed to hold more generally. This
is the reason for defining efficiency up to polynomial scaling: since any
computer can simulate the operation of any other up to polynomial overheads,
all computers can solve the same problems efficiently. In
Section~\ref{sub:classical_p_and_np} below, we consider the most
important classical \emph{complexity classes} that arise from these coarse but
robust definitions of efficiency.

The careful reader will have noticed that we restricted our statement of the
Church-Turing hypothesis to \emph{classical} computers. It is widely believed
that classical computers \emph{cannot} efficiently simulate quantum systems.
Certainly, directly simulating Schr\"odinger's equation on a polynomially sized
classical computer is problematic since the Hilbert space is exponentially
large. On the other hand, if we had a quantum computer with which to do our
simulation, the state space of our computer would also be a Hilbert space and
we could imagine representing and evolving complex states of the system by
complex states and evolutions of the quantum computer. This reasoning leads to
the strong \emph{quantum} Church-Turing hypothesis: that any reasonable
quantum computer may be efficiently simulated by any other. With this
hypothesis in hand, we may proceed to develop a robust classification of
\emph{quantum} complexity classes, as in
Section~\ref{sub:quantum_bqp_and_qma}.


\subsection{Polynomial reductions and worst-case behavior} 
\label{sub:polynomial_efficiency_reduction_and_worst_case_behavior}

\emph{Reduction} is the most important tool in complexity theory.
A decision problem A reduces to another problem B if there is a
polynomial time algorithm which can transform instances of A into
instances of B such that Yes-instances (No-instances) of A map to
Yes-instances (No-instances) of B. In this case, B is at least as
hard as A: any algorithm which could efficiently decide B would
be able to efficiently decide A as well. Just use the
transformation to convert the given instance of A into an
instance of B and then apply the efficient algorithm for B.

Reductions formalize the interrelationships between problems and
allow us to show that new problems are actually part of known
classes. Obviously, if we can reduce a problem A to a problem B
that we know how to solve efficiently, we have just shown how to
solve A efficiently as well. Conversely, if we have a problem C
which we believe is \emph{intractable} -- that is, not solvable by an
efficient algorithm -- and we can reduce it to another 
problem D, that suggests D should also be intractable. Using this
logic we can try to show that all kinds of interesting problems
ought to be intractable if we can find one to start with.


\subsection{Classical: P and NP} 
\label{sub:classical_p_and_np}

The most important complexity class is known as P -- this is the class of
decision problems which a classical computer can decide efficiently. More
precisely, a decision problem is in P if there exists an algorithm that runs
in polynomial time as a function of the input size of the instance and outputs
Yes or No depending on whether the instance is a Yes-instance or No-instance
of the problem. From a logical point of view, to show that a given problem is
in P we need to provide an efficient procedure to decide arbitrary instances.
Colloquially, P is the class of problems that are easy to solve.

We have already discussed one example, the DIVIDES problem, for which long
division constitutes a polynomial time decision procedure. Another example is
given by the energy evaluation problem: ``Does a specific configuration
$\sigma$ of a classical Ising Hamiltonian $H = \sum J_{ij} \sigma_i \sigma_j$
have energy less than $E$?" Here the instance is specified by a configuration
made of $N$ bits, a Hamiltonian function with  $N^2$ coupling terms and a
threshold energy $E$ (where all real numbers are specified with some fixed
precision). Since we can evaluate the energy $H(\sigma)$ using of order $N^2$
multiplications and additions and then compare it to $E$, this problem is also
in P.

The second most important complexity class is NP: this is the class of
decision problems for which there exists a scheme by which Yes-instances may
be efficiently verified by a classical computation. We may think of this
definition as a game between a prover and a verifier in which the prover
attempts, by hook or by crook, to convince the verifier that a given instance
is a Yes-instance. The prover provides the verifier with a proof of this claim
which the verifier can efficiently check and either Accept or Reject. NP
places no restrictions on the power of the prover -- only that Yes-instances
must have Acceptable proofs, No-instances must not have Acceptable proofs and
that the verifier can decide the Acceptability of the proof efficiently. We
note that there is an intrinsic asymmetry in the definition of NP: we do not
need to be able to verify that a No-instance is a No-instance.

For example, the problem ``Does the ground state of the Hamiltonian $H = \sum J_{ij}
\sigma_i \sigma_j$ have energy less than $E$?'' has such an efficient verification
scheme. If the prover wishes to show that a given $H$ has such low energy states, he
can prove it to the verifier by providing some configuration $\sigma$ which he claims
has energy less than $E$. The skeptical verifier may efficiently evaluate $H(\sigma)$
using the energy evaluation algorithm outlined above and if indeed $H(\sigma) < E$,
the skeptic would Accept the proof. If $H$ did not have such low energy states, then
no matter what the prover tried, he would be unable to convince the verifier to
Accept.

At first brush, NP seems a rather odd class -- why should we be
so interested in problems whose Yes-instances may be efficiently
checked? Of course, any problem we can decide efficiently (in P)
can be checked efficiently (because we can simply decide it!).
What of problems outside of NP? These do not admit efficient
verification schemes and thus certainly cannot have efficient
decision procedures. Moreover, even if, by some supernatural act
of intuition (not unusual in theoretical physics), we guess the
correct answer to such a problem, we would not be able to
convince anybody else that we were correct. There would be no
efficiently verifiable proof! Thus, NP is the class of problems
that we could ever hope to be convinced about.

Since 1971, the outstanding question in complexity theory (worth
a million dollars since the new millennium), has been ``Is P =
NP?'' This would be an astonishing result: it would state that all
of the difficulty and creativity required to come up with the
solution to a tough problem could be automated by a general
purpose algorithm running on a computer. The determination of the
truth of theorems would reduce to the simple matter of asking
your laptop 
to think about it. Since most scientists believe that there are
hard problems, beyond the capability of general purpose algorithms,
the consensus holds that P $\ne$ NP.%
\footnote{This may of course be
the bias of the scientists who don't want to be replaced by
omniscient laptops.}


\subsection{Quantum: BQP and QMA} 
\label{sub:quantum_bqp_and_qma}

The most important quantum complexity class is BQP -- this is the
class of decision problems which a quantum computer can decide
efficiently with bounded error (the B in the acronym). Since general
quantum algorithms have intrinsically stochastic measurement
outcomes, we have no choice but to allow for some rate of false-positive and
false-negative outcomes. As long as these rates are bounded
appropriately (say by 1/3), a few independent repetitions of the quantum
computation will exponentially suppress the probability of
determining the incorrect result. Thus, BQP is the quantum
analogue of P and plays a similar role in the classification of
decision problems. Since a quantum computer can simulate any
classical computation, P is contained in BQP.%
\footnote{For the expert, we note that a closer analogue of BQP is
  BPP, the class of decision problems which can be efficiently decided
  by a randomized classical algorithm with bounded error. In an
  attempt to minimize the onslaught of three letter acronyms, we have
  left this complication out.} 

The most important example of a BQP problem that is not known to be in P is
integer factoring. As a decision problem, this asks ``Given $N$ and $M$, does
the integer $N$ have a factor $p$ with $1 < p \le M$?'' In the 90's, Peter
Shor famously proved that factoring is in BQP by developing a quantum
factoring algorithm. There is no proof that factoring is classically hard 
(outside of P) -- nonetheless, many of the cryptography schemes on which
society relies for secure communication over the internet are only secure if
it is. Shor's algorithm renders all of these schemes useless if a large scale
quantum computer is ever built.

The quantum analogue of NP is the class QMA, Quantum Merlin-Arthur, which is
the class of decision problems whose Yes-instances can be efficiently checked
by a quantum computer given a quantum state as a proof (or witness). The
colorful name comes from the description of this class in terms of a game:
Merlin, all-powerful but less than trustworthy, wishes to prove to Arthur, a
fallible but well-intentioned individual who happens to have access to a
quantum computer, that a 
particular instance of a problem is a Yes-instance. Merlin, using whatever
supernatural powers he enjoys, provides Arthur with a quantum state designed
to convince Arthur of this claim. Arthur then uses his quantum computer to
decide, with some bounded error rate (say 1/3), whether to accept or reject
the proof. 

There are three primary differences between NP and QMA: 1) the verifier is a
quantum computer, 2) the proof is a quantum state, and, 3) the verification is
allowed a bounded error rate. The first two differences provide the class with
its additional quantum power; that the verifier is allowed a bounded error
rate is necessary due to quantum stochasticity, but not believed to be the
source of its additional power. We note that the particular error bound is
again somewhat arbitrary -- Arthur can exponentially improve the accuracy of a
noisy verification circuit by requesting Merlin provide him multiple copies of
the proof state and running his verifier multiple times
\cite{Aharonov:2002p4066}. Thus, even a verifier which falsely accepts
No-instances with probability up to $1/2 - 1/\mathrm{poly}(N)$ while accepting
valid proofs with probability $1/2$ only slightly larger can be turned into an
efficient bounded error QMA verifier through repetition.

An example of a QMA problem is given by the $k$-LOCAL HAMILTONIAN
problem:
       
\begin{description}
	\item[\bf Input:] A quantum Hamiltonian $H = \sum_m A_m$ composed of $M$
          bounded operators, each acting on $k$ qubits of an $N$ qubit Hilbert
          space. Also, two energies $a < b$, separated by at worst a
          polynomially small gap $b- a > 1/\textrm{poly}(N)$.
       
  \item[\bf Output:] Does $H$ have an energy level below $a$?
       
  \item[\bf Promise:] Either $H$ has an energy level below $a$ or all
          states have energies above $b$.
\end{description}

Here we have introduced the notion of a `promise' in a
decision problem. Promises are a new feature in our discussion: they impose a
restriction on the instances that a questioner is allowed to
present to a decision procedure. The restriction arises because the
algorithms and verification procedures we use to treat
promise problems need not be correct when presented with
instances that do not satisfy the promise -- an efficient
solver for LOCAL HAMILTONIAN could in fact fail on
Hamiltonians with ground state energies in the \emph{promise gap}
between $a$ and $b$ and we would still consider LOCAL
HAMILTONIAN solved. 

Heuristically, it is clear why we need the promise gap
for LOCAL HAMILTONIAN to be QMA: suppose we had a quantum
verifier which took a quantum state $| \psi \rangle$ and
tried to measure its energy $\epsilon = \langle \psi | H |
\psi \rangle$ through a procedure taking time $T$.
Time-energy uncertainty suggests that we should not be able
to resolve $\epsilon$ to better than $1/T$. Thus, if $T$ is
to be at most polynomially large in $N$, the verifier would
not be able to determine whether an $\epsilon$ exponentially
close to $a$ is above or below the threshold.

The actual construction of a verification circuit for the
LOCAL HAMILTONIAN problem is somewhat more subtle than simply
`measuring' the energy of a given state. As we will provide a very closely
related construction for the QSAT problem below, we do not include the verifier
for LOCAL HAMILTONIAN in these notes and instead refer the interested reader to
Ref.~\onlinecite{Aharonov:2002p4066}.  

The quantum analogue of the classical claim that P $\ne$ NP is that
BQP $\ne$ QMA -- a conjecture that is strongly believed for many of
the same reasons as in the classical case. 


\subsection{NP-Completeness: Cook-Levin} 
\label{sub:np_completeness_cook_levin}

In the early 1970s, Cook and Levin independently realized that
there are NP problems whose solution captures the difficulty of
the entire class NP. These are the so-called \emph{NP-complete}
problems. What does this mean?

A problem is NP-complete if it is both a) in NP (efficiently verifiable) and
b) any problem in NP can be reduced to it efficiently. Thus, if we had an
algorithm to solve an NP-complete problem efficiently, we could solve any
problem whatsoever in NP efficiently. This would prove P = NP with all of the
unexpected consequences this entails. Assuming on the contrary that P $\ne$
NP, any problem which is NP-complete must be intractable.

Let us sketch a proof of the Cook-Levin theorem showing the
existence of NP-complete problems. In particular, we will show
that classical 3-satisfiability, 3-SAT, is NP-complete. 3-SAT is
the decision problem which asks whether a given Boolean
expression composed of the conjunction of clauses, each involving
at most 3 binary variables, has a satisfying assignment.
Re-expressed as an optimization problem, 3-SAT asks, ``Does the
energy function 
\begin{equation}
	\label{eq:3sat_energy}
	H = \sum_m E_m(\sigma_{m_1}, \sigma_{m_2},	\sigma_{m_3}),
\end{equation}
acting on $N$ binary variables $\sigma_i$ in
which each local energy term $E_m$ takes values 0 or 1, have a zero
energy (satisfying) ground state?''%
\footnote{The interactions $E_m$ in 3-SAT are usually defined to penalize
  exactly one of the $2^3=8$ possible configurations of its input variables --
  but allow each of the $\binom{N}{3}$ possible 3-body interactions to appear
  multiple times in the sum. Thus, our definition is equivalent up to absorbing
  these terms together, which modifies the excited state spectrum but not the
  counting of zero energy satisfying states.} 

\subsubsection{3-SAT is in NP} 
\label{ssub:3_sat_is_in_np}

First, it is clear that 3-SAT is itself efficiently verifiable and therefore
in NP. If a prover wishes to prove that a particular instance $H$ is
satisfiable, she could provide a verifier a zero energy configuration. The
verifier would take this configuration and evaluate its energy (using
arithmetic in a polynomial number of steps) and thus be able to check the
validity 
of the claim. If $H$ is satisfiable, such a configuration exists. On the other
hand, if $H$ is not satisfiable, the prover would not be able to convince the
verifier otherwise because all states would have energy greater than zero.


\subsubsection{3-SAT is NP-complete} 
\label{ssub:3_sat_is_np_complete}

The tricky part is to show that 3-SAT is as hard as the entire
class NP. We need to show that \emph{any} possible NP problem can
be reduced to a 3-SAT problem by a polynomial transformation.
What is the only thing that all NP problems have in common? By
definition, they all have polynomial size verification procedures
which take as input a proposed proof that an instance is a
Yes-instance and output either Accept or Reject based on whether
the proof is valid. This verification procedure is what we will use
to provide the reduction to 3-SAT.

\begin{figure}[tbp]
	\centering
		\includegraphics[width=0.9\textwidth]{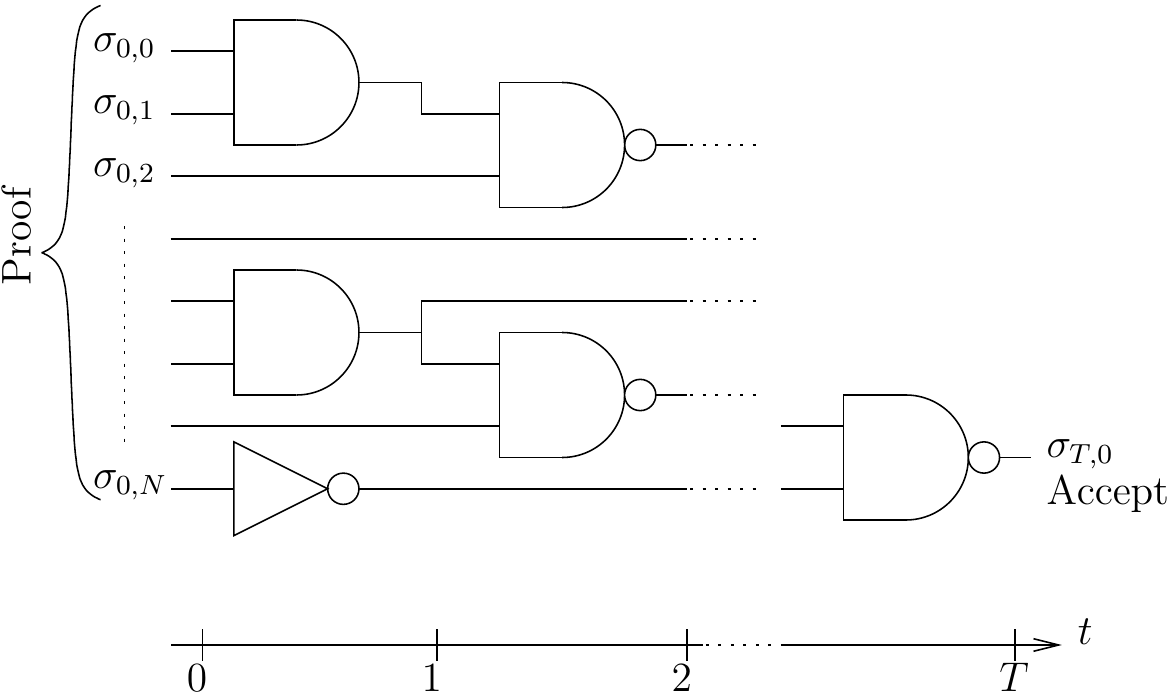}
	\caption{Circuit representing an NP verifier. The circuit
          depends on the particular instance and must be efficiently
          constructible by a polynomial time circuit drawing
          algorithm.} 
	\label{fig:figs_NP-verifier}
\end{figure}

Let us think of the verification procedure for a particular instance $A$ of
some NP problem as a polynomially sized Boolean circuit as in
Figure~\ref{fig:figs_NP-verifier}. The input wires encode the proposed proof
that $A$ is a Yes-instance and the output wire tells us whether to Accept or
Reject the proof. The gates in the figure are simply the usual Boolean logic
gates such as NAND and NOR, which take two input bits and provide one output
bit. Any Boolean circuit may be written using such binary operations with
arbitrary fan-out, so we assume that we can massage the verification circuit
into the form shown. Now we will construct an instance of 3-SAT encoding the
operation of this circuit. That is, if the instance is satisfiable, then there
exists a proof that the verifier accepts showing that the original NP problem
is a Yes-instance and conversely, if the instance is not satisfiable, then no
such proof exists and the original NP problem is a No-instance.

\begin{figure}[tbp]
	\centering
		\includegraphics{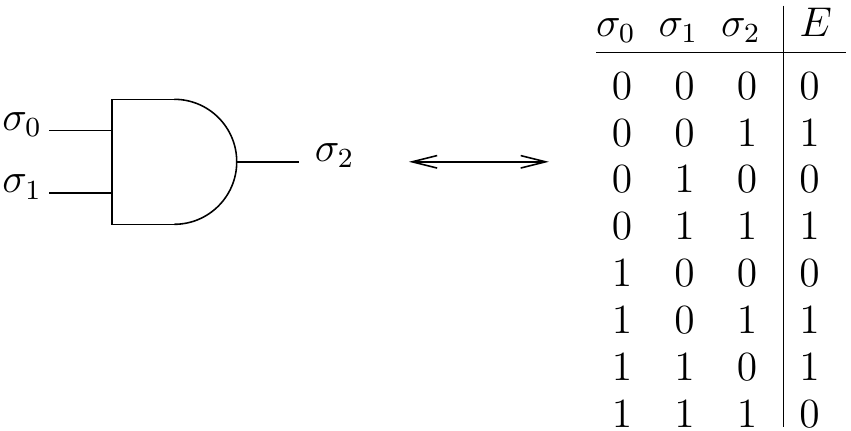}
	\caption{Interpretation of Boolean AND gate as three-body interaction.}
	\label{fig:figs_and-energy}
\end{figure}

The 3-SAT instance is very simple to construct if we simply change our point
of view on the picture in Figure~\ref{fig:figs_NP-verifier}. Instead of
viewing it as a Boolean circuit operating from left to right, let us view it
as the interaction graph for a collection of $O(N \times T)$ binary bond
variables -- one for each of the wires in the circuit: the input bits
of the proof, the output bit and each of the intermediate
variables. Each gate then 
specifies a 3-body interaction $E_m$ for the adjacent variables which we
define to take the value $0$ for configurations in which the variables are
consistent with the operation of the gate and $1$ otherwise. See
Figure~\ref{fig:figs_and-energy}. We now add a final 1-body term on the output
bit of the verification circuit which penalizes the Reject output. We now have
an Ising-like model with polynomially many 3-body interactions and non-negative
energy. 

That's it. If the 3-SAT instance described by
Figure~\ref{fig:figs_NP-verifier} has a zero energy ground state, then there
is a configuration of the wire variables such that the circuit operates
correctly and produces an Accept output. In this state, the input wires
represent a valid proof showing that the original instance was a Yes-instance.
On the other hand, if the 3-SAT instance is not satisfiable, no state exists
such that the circuit operates correctly and the output produced is always
REJECT. Thus we have shown that all problems in NP can be efficiently 
reduced to 3-SAT. 

Now that we have one problem, 3-SAT, which is NP-complete, it is
straightforward to show the existence of many other NP-complete problems: we
need only find reductions from 3-SAT to those problems. Indeed, a veritable
menagerie of NP-complete problems exists (see \emph{e.g.},
\cite{Garey:1979ud}) including such famous examples as the traveling salesman
problem and graph coloring. A more physics oriented example is that of
determining the ground state energy of the $\pm J$ Ising model in 3 or more
dimensions \cite{Barahona:1982p157}.



\subsection{QMA-Completeness: Kitaev} 
\label{sub:qma_completeness_kitaev}

The complexity class QMA provides the quantum analogue to NP and, just like
NP, it contains complete problems which capture the difficulty of the entire
class. Kitaev first introduced the QMA-complete problem 5-LOCAL HAMILTONIAN in
the early '00s and proved its completeness using a beautiful idea due to
Feynman: that of the history state, a superposition over computational
histories. The quantum Cook-Levin proofs are 
somewhat more complicated than the classical case and we will only sketch them
here (see \cite{Bravyi:2006p4315} for more details). For simplicity and to
connect with the statistical study 
undertaken in the later sections, we restrict our attention to the
slightly simpler problem of $k$-QSAT, which is QMA$_1$-complete for $k \ge 4$.
QMA$_1$ is the variant of QMA in which the verification error is one-sided:
Yes-instances may be verified with no errors while invalid proofs still
occasionally get incorrectly accepted.

First, let us define $k$-QSAT a bit more carefully:

\begin{description}
	\item[\bf Input:] A quantum Hamiltonian $H = \sum_m \Pi_m$ composed of $M$
      projectors, each acting on at most $k$ qubits of an $N$ qubit Hilbert space. 

  \item[\bf Promise:] Either $H$ has a zero energy state or all states
    have energy above a promise gap energy $\Delta > 1
    /\textrm{poly}(N)$. 

  \item[\bf Question:] Does $H$ have a zero energy ground state?
\end{description}

Now, we sketch the proof that QSAT is QMA$_1$-complete.

\subsubsection{QSAT is QMA$_1$} 
\label{ssub:qsat_is_qma_1}

To show that QSAT is QMA$_1$, we need to find an efficient quantum
verification scheme such that (a) there exist proofs for Yes-instances which
our verifier always accepts and (b) any proposed proof for a No-instance will
be rejected with probability at least $\epsilon = 1/\textrm{poly}(N)$. This
rather weak requirement on the bare false-Acceptance rate can be bootstrapped
into an arbitrarily accurate verification scheme by repetition, as sketched in
Section~\ref{sub:quantum_bqp_and_qma} above. 

Given an instance $H = \sum_m \Pi_m$, the obvious proof  is for Merlin to
provide a state $| \Psi \rangle$ which he alleges is a zero energy
state. Arthur's verification procedure will be to check this
claim. The verifier works by measuring each of the $\Pi$ in some 
pre-specified order on the state. That this can be done efficiently follows
from the fact that $\Pi$ acts on no more than $k$ qubits and therefore its
measurement can be encoded in an $N$-independent number of quantum gates.
Clearly, if $| \Psi \rangle$ is a zero-energy state, it is a zero-energy
eigenstate of each of the $\Pi$ and therefore all of these measurements will
produce $0$ and the verifier accepts. This checks condition (a) above
and we say our verification scheme is complete.%
\footnote{The feature that one can do these measurements by local operations
  and that they provide probability 1 verification of ground states is a
  special feature of the QSAT Hamiltonian which allows it somewhat to evade the
  heuristic expectations of time-energy uncertainty.} 

On the other hand, if $H$ is a No-instance, it has a ground state energy above
the promise gap $\Delta$ and $| \Psi \rangle$ necessarily has overlap with
the positive eigenspaces of at least some of the $\Pi$. It is a short
computation to show that the probability that all of the measurements return
$0$ will then be bounded above by $1 - \Delta/N^k \sim 1 - 1/\textrm{poly}(N)
$. Thus, No-instances will be rejected with probability at least
$\epsilon = 1 / \textrm{poly}(N)$ and our verification scheme is
sound. 


\subsubsection{QSAT is QMA$_1$-complete} 
\label{ssub:qsat_is_qma_1_complete}

 Just as in the classical Cook-Levin proof, we need to show that \emph{any}
QMA$_1$ problem can be reduced to solving an instance of QSAT. We again
exploit the only thing that all QMA$_1$ problems have in common: their quantum
verification algorithm. We will take the quantum circuit representing this
verifier and construct from it a QSAT Hamiltonian whose ground state energy
is directly related to the maximal acceptance probability of the
verification circuit.

\begin{figure}[tbp]
	\centering
		\includegraphics[width=0.9\textwidth]{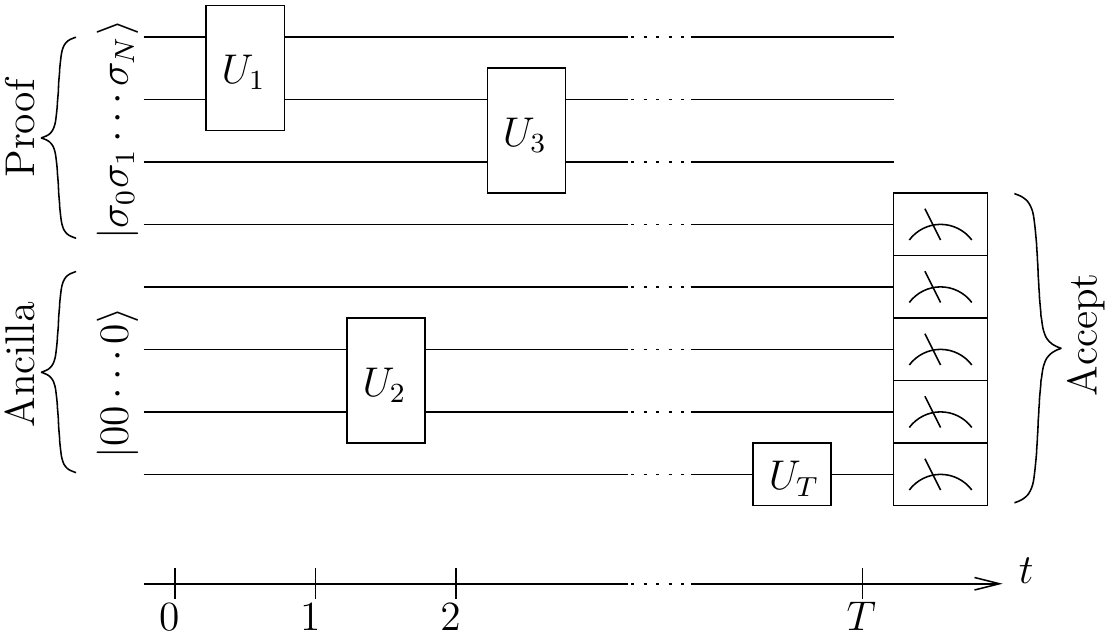}
	\caption{QMA verification circuit. The circuit depends on the
          instance and must be constructible by an efficient algorithm
          given the instance.  We have drawn the circuit so that there
          is a single local gate per time step, so $T =
          \textrm{poly}(N)$.} 
	\label{fig:figs_QMA-verifier}
\end{figure}

 Let $A$ be an arbitrary instance of a QMA$_1$ problem $L$. Then $A$ has a
polynomial sized quantum verification circuit as in
Figure~\ref{fig:figs_QMA-verifier}. This circuit takes as input a quantum
state encoding a proof that $A$ is a Yes-instance of $L$, along with some
ancilla work qubits in a fiducial $0$ state, then performs a sequence of $T$
unitary one and two qubit gates and finally measures the output state of some
subset of the qubits. If $A$ is a Yes-instance, then there exists a valid
input state such that all of the output bits will yield $0$ with probability
1. Otherwise, at least one of the output bits will read $1$ with probability
polynomially bounded away from 0 for any input state $| \sigma_0 \cdots
\sigma_N \rangle$.%
\footnote{The observant reader will notice the addition ancillae qubits. These
  are necessary when computation is done by reversible gates, as in unitary
  circuit computation. We leave it as an exercise to figure out
  why the absence of ancillae would make the verification circuit unsound.} 

 We now construct a single particle hopping Hamiltonian whose ground state
encodes the operation of this quantum circuit. We introduce a \emph{clock}
particle hopping on a chain of length $T$ and label its position states by $|
t \rangle$. We endow this particle with an enormous `spin': a full $N$ qubit
Hilbert space (dimension $2^N$). As the particle hops, `spin-orbit' coupling induces rotations
on the $N$ qubit space which correspond to the unitary gates in the
verification circuit. To wit:
\begin{equation*}
 H_p = \frac{1}{2} \sum^{T-1}_{t=0}  \left(- |t+1 \rangle \langle t| \otimes U_{t+1} -
\ket{t}\bra{t+1}\otimes U^\dagger_{t+1} + |t \rangle \langle t | + |t+1 \rangle \langle t+1|\right)
\end{equation*}

 The terms of this Hamiltonian have been normalized and shifted such that each
is a projector with energies $0$ and $1$, but otherwise it is just a 1-D
hopping problem with Neumann boundary conditions. \footnote{In fact, Neumann boundary conditions (which stipulate the value of the derivative of the solution to a differential equation) apply to the problem obtained in the time continuum limit of the discrete hopping Hamiltonian under consideration.} Indeed, there is a simple
basis transformation in which the spin-orbit coupling disappears entirely.
This consists of rotating the basis of the position $t$ spin space by a sequence of unitary transformations
$U^\dagger_1 U^\dagger_2 \cdots U^\dagger_t$. 

In this representation, we see
that the $2^N$ spin components decouple and the system is really $2^N$
copies of the Neumann chain. Thus, the spectrum is $(1 - \cos k)/2$, a cosine dispersion with
bandwidth $1$, ground state energy $0$ at wave vector $k = 0$ and allowed $k= n
\pi / (T+1)$. 

 The propagation Hamiltonian has (zero energy) ground states of the form (in
the original basis):
\begin{equation}
 |\psi\rangle = \frac{1}{\sqrt{T+1}}\sum_t | t \rangle \otimes U_t U_{t-1} \cdots U_1 | \xi
\rangle
\end{equation}
where $| \xi \rangle$ is an arbitrary `input state' for the $N$ qubit space.
This state $| \psi \rangle$ is called the \emph{history state} of the computation performed by
the verification circuit given input $|\xi\rangle$. It is a sum over the state
of the quantum computation at each step in the circuit evolution. Any correct
computation corresponds to a zero energy history state -- incorrect
computations will have a non-zero overlap with higher energy hopping states.

 Now, we have a Hamiltonian that encodes the correct operation of the
verification circuit. We simply need to add terms that will penalize
computations which do not correspond to appropriate input states and accepting
output states. These terms affect the computational state at times $t=0$ and
$T$, so they are boundary fields from the point of view of the hopping
problem. In general, they should split the $2^N$ degeneracy of the pure
hopping in $H_p$, and since they are positive operators, lift the ground state
energy.

The initialization term is simply a sum over the ancilla qubits of projectors
penalizing $| \xi \rangle$ with incorrectly zeroed ancillae:
\begin{equation}
	 H_i = \sum_{j\in \textrm{Ancilla}} |0 \rangle \langle 0 |_t \otimes | 1 \rangle
	\langle 1 |_j	
\end{equation}
 Similarly, the output term penalizes output states which overlap $| 1
\rangle$ on the measured output bits:
\begin{equation}
	 H_o = \sum_{j\in \textrm{Accept}} |T \rangle \langle T|_t \otimes | 1 \rangle
	\langle 1|_j	
\end{equation}

 Now we consider the full Hamiltonian
\begin{equation}
 H = H_i + H_p + H_o	
\end{equation}
If $| \psi \rangle$ is a zero energy state of $H$, then it is a zero energy
state of each of the three pieces. Hence, it will be a history state in which
the input state $|\xi\rangle$ has appropriately zeroed ancillae and the output
state has no overlap with $|1\rangle$ on the measured qubits -- thus,
$|\xi\rangle$ is a proof that the verifier accepts with probability 1 and the
original instance $A$ is a Yes-instance. Conversely, if such a proof state $|
\xi \rangle$ exists then the history state built from it will have zero
energy.

 It is somewhat more work to show the soundness of the construction: $A$ is a
No-instance if and only if the Hamiltonian $H$ has ground state energy bounded
polynomially away from 0 \cite{Aharonov:2002p4066}. The intuition is
straightforward -- the strength of the boundary fields in the
basis transformed hopping problem for a given spin state corresponds to the
acceptance probability of the associated input state. 
Since these repulsive fields lift Neumann conditions, they raise the ground state energy quadratically in 1/T -- they effectively force the ground state wavefunction to bend on the scale of T. For a No-instance, since no spin sector is both valid and accepting, all states must gain this inverse quadratic energy.%
\footnote{This is overly simplified: in the absence of the output term $H_o$,
the gauge transformed problem can be thought of as $2^N$ decoupled hopping
chains, some fraction of which have boundary fields at $t=0$. The output term
is not simply a field on these chains -- it couples them and in principle
allows hopping between them as a star of chains. The upshot is that the
repulsive (diagonal) piece outweighs the off-diagonal mixing.}

To be a bit more precise, we assume for contradiction that we have a state
$\ket{\psi}$ with energy exponentially small in $N$ (hence smaller than any
polynomial in $N$ or $T$): 
\begin{equation}
	\bra{\psi}H\ket{\psi} = \bra{\psi}H_i\ket{\psi} + \bra{\psi}H_p\ket{\psi} +
    \bra{\psi}H_o\ket{\psi} \le O(e^{-N}) 
\end{equation}
Since each term is positive, each is bounded by the exponential. The hopping
Hamiltonian $H_p$ has a gap of order $1/T^2$ for chains of length $T$, thus if
we decompose $\ket{\psi}$ into a zero energy piece (a history state) and an
orthogonal complement, 
\begin{equation}
	\ket{\psi} = \frac{\sqrt{1 - \alpha^2}}{\sqrt{T+1}}\sum_t
        \ket{t} \otimes U_t\cdots U_1 \ket{\xi} + \alpha \ket{\mathrm{Exc}} 
\end{equation}
we must have exponentially small overlap onto the complement:
\begin{equation}
	O(e^{-N}) > \bra{\psi}H_p\ket{\psi} = \alpha^2
        \bra{\mathrm{Exc}}H_p\ket{\mathrm{Exc}} > \alpha^2 O(1/T^2) 
\end{equation}
In other words,
\begin{equation}
	\ket{\psi} = \frac{1}{\sqrt{T+1}}\sum_t \ket{t} U_t\cdots U_1 \ket{\xi} + O(e^{-N})
\end{equation}
The input term $H_i$ has energy 0 for valid input states $\ket{\xi^V}$
and energy at least 1 for invalid states $\ket{\xi^I}$. Thus,
decomposing $\ket{\xi} = \sqrt{1-\beta^2} \ket{\xi^V} + \beta
\ket{\xi^I}$, we find (abusing notation and dropping explicit
reference to the $\ket{t=0}$ sector on which $H_i$ acts): 
\begin{eqnarray}
	O(e^{-N}) > \bra{\psi}H_i\ket{\psi} = \frac{\beta^2}{T+1}
        \bra{\xi^I}H_i \ket{\xi^I} + O (e^{-N}) \ge
        \frac{\beta^2}{T+1} + O(e^{-N}) 
\end{eqnarray}
In other words,
\begin{equation}
	\ket{\psi} = \frac{1}{\sqrt{T+1}}\sum_t \ket{t} U_t\cdots U_1 \ket{\xi^V} + O(e^{-N})	
\end{equation}
Finally, considering the output term we find:
\begin{eqnarray}
	\bra{\psi} H_o \ket{\psi} &=& \frac{1}{T+1}
        \bra{\xi^V}U^\dagger_1\cdots U^\dagger_t H_o U_t \cdots U_1
        \ket{\xi^V} + O(e^{-N}) \nonumber \\ 
	&=& \frac{p_r}{T+1} + O(e^{-N})
\end{eqnarray}
where $p_r$ is the probability that the original QMA$_1$ verifier
rejects the proposed proof $\ket{\xi^V}$ for the No-instance $A$.
Since this rejection probability is bounded below by a constant, the
state $\ket{\Psi}$ cannot possibly have exponentially small energy.  

We have reduced the arbitrary QMA$_1$ instance $A$ to asking about the zero
energy states of a hopping Hamiltonian $H$ constructed out of projectors. 
This is almost what we want. We have a Hamiltonian constructed out of
a sum of projectors but they each act on three qubits tensored with a
(large) particle hopping space rather than on a small collection of
qubits.  

The final step in the reduction to $k$-QSAT is to represent the single
particle Hilbert space in terms of a single excitation space for a chain of
clock qubits in such a way that we guarantee the single particle sector is
described by $H$ above and that it remains the low energy sector. We refer the
interested reader to the literature for more details on these
clock constructions. Each of the projectors of $H$ becomes a joint projector
on one or two of the computational (spin) qubits and some number of the clock
qubits (two in \cite{Bravyi:2006p4315}). The final 4-QSAT Hamiltonian will then be given
by a sum of projectors involving at most 4 qubits \begin{equation} H = H_i +
H_p + H_o + H_c \end{equation} where $H_c$ acts on the clock qubits to
penalize states which have more than one clock particle. This concludes our brief overview of complexity theory. We next turn to a review of results obtained by applying ideas from (quantum) statistical mechanics to random ensembles of classical and quantum $k$-SAT.




\section{Physics for complexity theory} 
\label{sec:physics_for_complexity_theory}

There are two main ways for physicists to contribute to complexity theory. One is to
bring to bear their methods to answer some of the questions posed by complexity
theorists. Another is to introduce concepts from physics to ask new types of
questions, thereby providing an alternative angle, permitting a broader view and new
insights. This section is devoted to the illustration of this point, using the $k$-SAT
problem introduced above as a case in point. In particular, we discuss both classical
$k$-SAT and its quantum generalisation $k$-QSAT
\cite{Laumann:2010p7275,Bravyi:2006p4315}.

\subsection{Typical versus worst-case complexity} 
\label{sub:typical_versus_worst_case_complexity}

As explained above $k$-SAT is NP complete for $k \geq 3$. Thus, for any given
algorithm, we expect that there are instances which will take an exponentially
long time to solve. However, we ought not be too discouraged -- some instances
of $k$-SAT may be parametrically easier to solve than others, and these may be
the ones of interest in a given context. To make this more precise, it is
useful to introduce the concept of typical, as opposed to worst-case,
complexity.

In order to define typicality, one can consider an ensemble in which each
problem instance is associated with a probability of actually occurring. Typical
quantities are then given by stochastic statements, e.g. about a median time
required for solving problem instances, which may differ substantially from
the corresponding average, or indeed the worst-case, quantities when the latter have a sufficiently
small weight in the ensemble. Precisely what quantities to calculate depends
on the aspects of interest. For instance, a median run-time is not
much affected by a small fraction of exponentially long runs, while these may
dominate the expectation value of the run-time.

It is worth emphasizing again that the polynomial
reductions discussed in Section~\ref{sec:complexity_theory_for_physicists}
provide a characterization of the \emph{worst case} difficulty of solving
problems. The reductions and algorithms in this context must work for
\emph{all} instances of a problem. Reductions however may transform
typical instances of A into rather `atypical' instances of B. Whether a useful
framework of 
reductions can be defined that preserve typicality is an open question (see
Chapter 22 of Ref. \cite{Arora:2009zv}), but the study of typical instances of
particular hard problems has itself been a fruitful activity, as we will
discuss in the following.


\subsection{Classical statistical mechanics of $k$-SAT} 
\label{sub:classical_statistical_mechanics_of_k_sat}

We now give an account of an analysis of such an ensemble for classical
$k$-SAT. For completeness, let us begin with reviewing the original definition
of classical $k$-SAT, expanding on the brief definition provided in
Section~\ref{sub:np_completeness_cook_levin}. Indeed, the original
computer science definition of satisfiability looks somewhat different from the
Hamiltonian problem we introduced.  
Consider a set of
$N$ Boolean variables $\{x_i \mid i = 1 \dots N\}$, i.e. each variable ${\rm
  x}_i$ can take two values, true or false (in which cases the negation 
$\bar{\rm x}_i$ is false or true, respectively). Classical $k$-SAT
asks the question, ``Does the Boolean expression:  
\begin{equation}
{\cal{C}} = \bigwedge\limits^{\rm M}_{m=1} {\rm C}_{m}
\label{Eq:1}
\end{equation}
evaluate to true for some assignment of the ${\rm x}_i$?'' Here, each
clause is composed of a disjunction of $k$ literals, e.g. for $k = 3$:
\begin{equation}
{\rm C}_{m} = {\rm x}_{m_1} \vee\bar{\rm
  x}_{m_2} \vee{\rm x}_{m_3}   
\label{Eq:2}
\end{equation}
where each variable occurs either affirmed (${\rm x}_{m_1}$) or
negated ($\bar{\rm x}_{m_2}$). Hence, there are 2$^k$ possible clauses for
a given $k$-tuplet $\{{\rm x}_{m_j} \mid j = 1 \dots k\}$.%
\footnote{The symbols $\wedge$ and $\vee$ are the
  Boolean operators `and' and `or'.} 

This definition is equivalent to the definition in terms of the $k$-body
interacting spin Hamiltonian of Eq.~\eqref{eq:3sat_energy}. To obtain
a spin Hamiltonian from the collection of clauses, Eq.~\eqref{Eq:1},
we 
convert the Boolean variables ${\rm x}_i$ into Ising spins $\sigma_i = \pm 1$,
with $\sigma_i = +1 (-1)$ representing ${\rm x}_i$ being true (false). A clause
then becomes a $k$-spin interaction designed such that the satisfying
assignments evaluate to energy 0, and the forbidden assignment to energy 1. For
instance, the clause given in Eq.~\eqref{Eq:2} turns into:
\begin{equation}
\HH_{m} = 2^{-3} \left( 1 - \sigma_{i_{m_1}}
\right) \left( 1 + \sigma_{i_{m_2}} \right) \left( 1 -
  \sigma_{i_{m_3}} \right) ~~~.
\label{Eq:halpha}
\end{equation}

\noindent (It is this formulation of classical $k$-SAT that will lend itself
naturally to a quantum generalisation, which we describe below.)

The $k$-SAT ensemble is now random in two ways:
\begin{description}
\item[(R1)] each $k$-tuple occurs in $H$ with probability $p = \alpha N
  \left/ \left( N \atop k \right) \right.$
\item[(R2)] each $k$-tuple occurring in $H$ is randomly assigned one of
the $2^k$ possible clauses. 
\end{description}

\begin{figure}[tbp]
	\centering
		\includegraphics[width=0.9\textwidth]{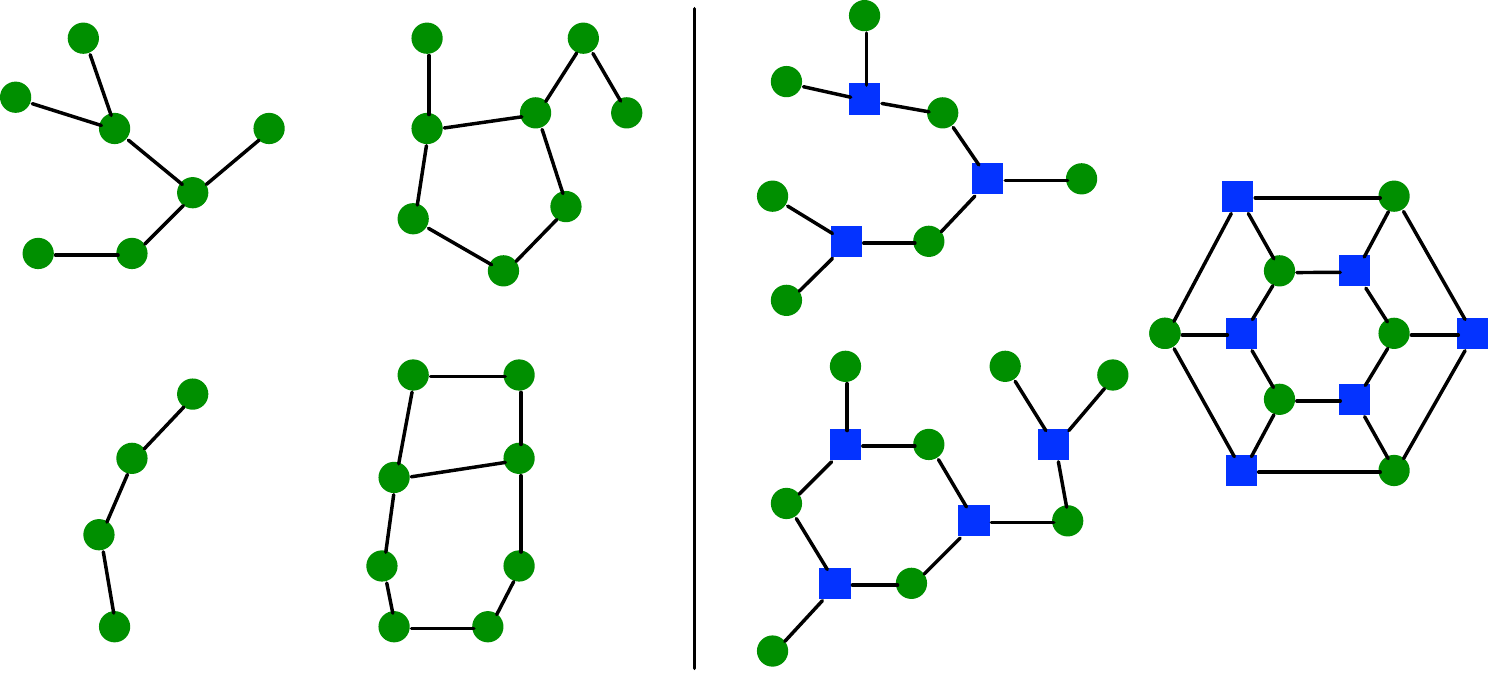}
	\caption{Examples of random interaction graphs for (a) 2-SAT and (b)
3-SAT, respectively. The (green) circles represent qubits. (a) The clusters, clockwise from bottom left,
are chain, tree, clusters with one and two closed loops (``figure 
eight''). The short closed loops, as well as the planarity of the
graphs, are not representative of the large-$N$ limit. (b) Each (blue) square
represents a clause connected to 3 nodes. Clockwise from top
left are a tree, a graph with nontrivial core and a graph with simple loops but
no core.} 
	\label{fig:figs_graphs}
\end{figure}

\noindent Here, we have introduced a parameter $\alpha$ for the number of
clauses, ${\rm M} = \alpha N$, which is proportional to the number of
variables%
\footnote{Actually, only the expectation value of ${\rm M}$ equals $\alpha N$.
The Poissonian distribution for ${\rm M}$ of course has vanishing relative
fluctuations $(\langle {\rm M}^2 \rangle - \langle {\rm M} \rangle^2)/\langle
M \rangle^2$ as $N \to \infty$.}
because there are $\left( N \atop k \right)$ possible $k$-tuples. The
`interactions' can be pictorially represented by an interaction graph,
Fig.~\ref{fig:figs_graphs}. This is a bipartite graph, one sublattice of
which has $N$ nodes, denoted by circles representing the ${\rm x}_i$, and the
${\rm M}$ nodes of the other sublattice denoted by triangles represent the
clauses ${\rm C}_{m}$. Each triangle is connected to the $k$ variables
participating in the clause it represents, whereas each variable is connected
to all clauses it participates in, which implies an average coordination of
$\alpha k$, with a Poissonian distribution. The random graph thus constructed
contains all the information on a given problem instance if we label each
triangle with which of the 2$^k$ possible clauses it represents. This graph
will be used for random quantum $k$-SAT as well, where the Boolean variables
and clauses will be replaced by appropriate quantum generalisations.

\subsection{Schematic phase diagram of classical random $k$-SAT} 
\label{sub:phase_diagram_of_classical_random_k_sat}

\begin{figure}[tbp]
	\centering
		\includegraphics[width=0.9\textwidth]{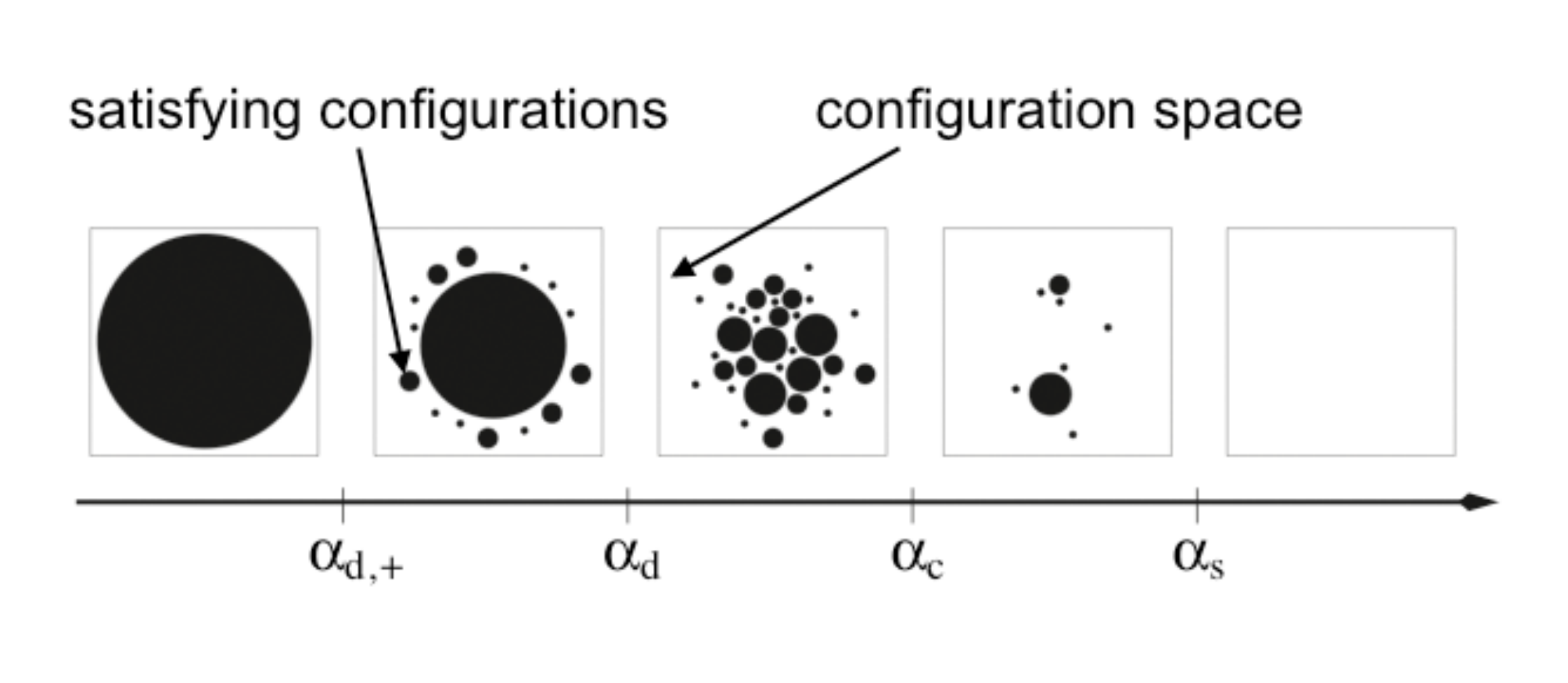}
	\caption{Schematic phase diagram for random classical $k$-SAT ($k\ge
      4$) \cite{Krzakala:2007p8722}. Actually, configuration space is very
      high-dimensional (an $N$-dimensional hypercube), and the cartoons are
      only suggestive of the actual structure of the space of solutions, for
      the real complexity of which our everyday intuition from low dimensions
      may be quite inadequate.} 
	\label{fig:figs_pnas-2007-ksat-pd}
\end{figure}

In Fig. \ref{fig:figs_pnas-2007-ksat-pd}, we show a schematic phase diagram
for random $k$-SAT. The first question one might ask is: is there a
well-defined phase transition, at some value $\alpha = \alpha_s (k)$, such
that instances for $\alpha < \alpha_s$ are satisfiable, and those for $\alpha
> \alpha_s$ are not? It has been shown that there exists such a transition for
the random ensemble. This does not mean that \emph{all} instances with $\alpha
< \alpha_s$ are satisfiable: given an UNSAT instance with $N$ sites and
$\alpha N$ clauses, one could simply add $N$ disconnected sites to get a new
UNSAT instance with $\alpha' = \alpha/2$. What is true instead is that the
probability of having such an UNSAT graph with $\alpha' < \alpha_{s}$ is
exponentially small in $N$, so that for $N \to \infty$, such graphs do not
arise with a probability approaching 1.

It is easy to provide a very rough estimate for where this happens, by
adapting an idea of Pauling's which, amusingly, was devised for estimating the
configurational entropy of the protons in water ice. We consider the clauses
as constraints, each of which `decimates' the number of allowed configurations
by a factor ($1-2^{-k}$): only 1 out of the 2$^k$ possible configurations of
variables of any given clause is ruled out. For ${\rm M} = \alpha N$ such
constraints, one is left with $2^N (1-2^{-k})^{\alpha N}$ solutions. In the
thermodynamic limit, this number vanishes for $\alpha > \alpha_{wb} = -1 /
\log_2 (1-2^{-k}) \sim 2^k\log2$. In the $k$-SAT literature, this is known as
the `first-moment bound', for which there is a straightforward rigorous
derivation. To find rigorous \emph{upper bounds} one should instead employ
different techniques, with inequalities coming from the analysis of the
\emph{second moment} of the number of solutions (after appropriately
restricting the ensemble to reduce the fluctuations) \cite{Dubois:2001p9880}.
It is interesting here to note that for large $k$ the upper bounds and lower
bounds converge, becoming a prediction for the actual location of the
threshold.

The SAT-UNSAT transition is not the only transition of this problem, though.
As indicated in Fig.~\ref{fig:figs_pnas-2007-ksat-pd}, statistical
mechanical methods imported from the study of spin glasses have been used to
establish finer structure in the SAT phase. This plot shows a set of cartoons
of configuration space, indicating the location of satisfying assignments. For
$N$ variables, configuration space is an $N$-dimensional hypercube and this
plot indicates, in a two-dimensional `projection', how `close' satisfying
assignments are to each other. Roughly, two solutions belong to the same
cluster if they can be reached via a sequence of satisfying configurations
such that two consecutive ones differ by $O(N^\beta)$ variables with $\beta <
1$ \cite{Krzakala:2007p8722}. 

Figure~\ref{fig:figs_pnas-2007-ksat-pd} thus documents a set of transitions in
the clustering of satisfying assignments. For the smallest $\alpha$, all
solutions belong to one single giant cluster -- the full hypercube for $\alpha
= 0$ -- and then there is a successive break-up into smaller, and increasingly
numerous clusters as $\alpha$ grows \cite{Krzakala:2007p8722}. 

This structure of configuration space should have ramifications for how hard
it is to solve the corresponding instances: small-scale clusters indicate a
rugged energy landscape, with numerous local minima providing opportunities
for search algorithms to get stuck. Indeed, all known algorithms slow down
near $\alpha_s$. That said, many simple approaches to random $3$-SAT
problems actually do quite well even in the clustered phases and the detailed
relationship between clustering in configuration space and algorithmic
difficulty is subtle, somewhat detail dependent and an ongoing
research topic. It is particularly worth noting that
even the simplest random greedy algorithms typically work across most
of these transitions, at least for $k$-SAT. Indeed, while the
identification of distinct phases has grown to give a phase diagram
replete with fine structure, the portion of the phase
diagram containing `hard' instances -- those for which deciding
satisfiability takes exponentially long typically -- has by now been
pushed back to a tiny sliver at $\alpha_s$ of width $\delta\alpha <
\alpha_s/100$. In the meantime, however, the action has started to
shift to other problem ensembles which at the time of writing have
proven more robustly difficult. 

The derivation of this phase diagram was obtained using methods imported from
the study of spin glasses, in particular the \emph{cavity method}
\cite{Krzakala:2007p8722,Hartmann:2005ys}. The insights thus gained have lead
to the development of an impressive arsenal of techniques for not only
determining whether or not a $k$-SAT problem instance is soluble, but also for
actually finding solutions in the form of satisfying assignments
\cite{Braunstein:2005p110,Mezard:2003bd,Mezard:2002p94,Mezard:2002p120}.  
In the following section, we provide a brief introduction to cavity analysis.


\subsection{Cavity analysis} 
\label{sub:cavity_analysis}

\newcommand{\bs}{\backslash}
The cavity method is a cluster of techniques and heuristics for solving
statistical models on sparse, tree-like interaction graphs $G$. In this
approach, one determines the behavior of the model on $G$ by first analyzing
the restriction of the model to so-called \emph{cavity graphs}. A cavity graph
$G\bs\set{i}$ is formed by carving site $i$ out of $G$:
\begin{equation*}
\includegraphics{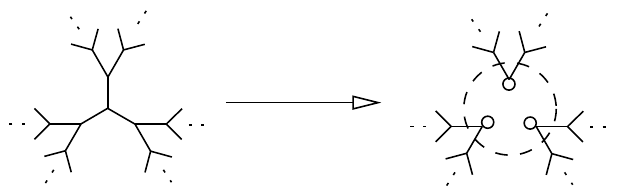}
\end{equation*}
The neighbors of $i$, $\partial i$, which now sit at the boundary of the
cavity in $G \bs \set{i}$, are called \emph{cavity spins}. The central
assumption of the cavity method is that cavity spins are statistically
independent in the absence of site $i$ because they sit in disconnected
components of $G\bs \set{i}$. This assumptions massively simplifies the
evaluation of observables in such models, ultimately leading to efficient
procedures for finding ground states, evaluating correlation functions and
determining thermodynamic free energies, phase diagrams, and clustering
phenomena in models with quenched disorder. 

The ensemble of interaction graphs $G$ that arise from the rule (R1) have
loops and thus do not fall into disconnected pieces when a cavity site $i$ is
removed. Nonetheless, for large $N$, any finite neighborhood of a randomly
chosen point in $G$ is a tree with high probability. That is to say, $G$ does
not contain short closed loops and we call it \emph{locally tree-like}. For
such $G$, we can hope that the cavity assumption will hold at least to a good
approximation. 

That neighborhoods in $G$ are trees can be seen as follows: the
subgraph consisting of site $i$ and its neighbors has on average $n_1 = (1 +
\alpha k)$ out of the $N$ sites. The $\alpha k$ neighbors will in turn have
$\alpha k$ further neighbors, so that the subgraph containing those as well
has approximately $n_2 = 1 + (\alpha k) + (\alpha k)^2$ sites. Up to the
$\gamma$-th nearest neighbors, the resulting subgraph grows exponentially,
$n_\gamma \sim (\alpha k)^\gamma$. Obviously, $n_\gamma \leq N$, so that closed
loops must appear at length $\gamma_c = (\ln N)/\ln(\alpha k)$. For $\gamma <
\gamma_c$, $n_\gamma \ll N$ due to the exponential growth of $n_\gamma$, so
that the randomly chosen interaction partners are overwhelmingly likely to be
drawn from the sites not yet included in the neighborhood. Thus, the length of loops on $G$ deverges with $N$, although excruciatingly (logarithmically) slowly.

\begin{figure}[tbp]
	\centering
		\includegraphics[width=2in]{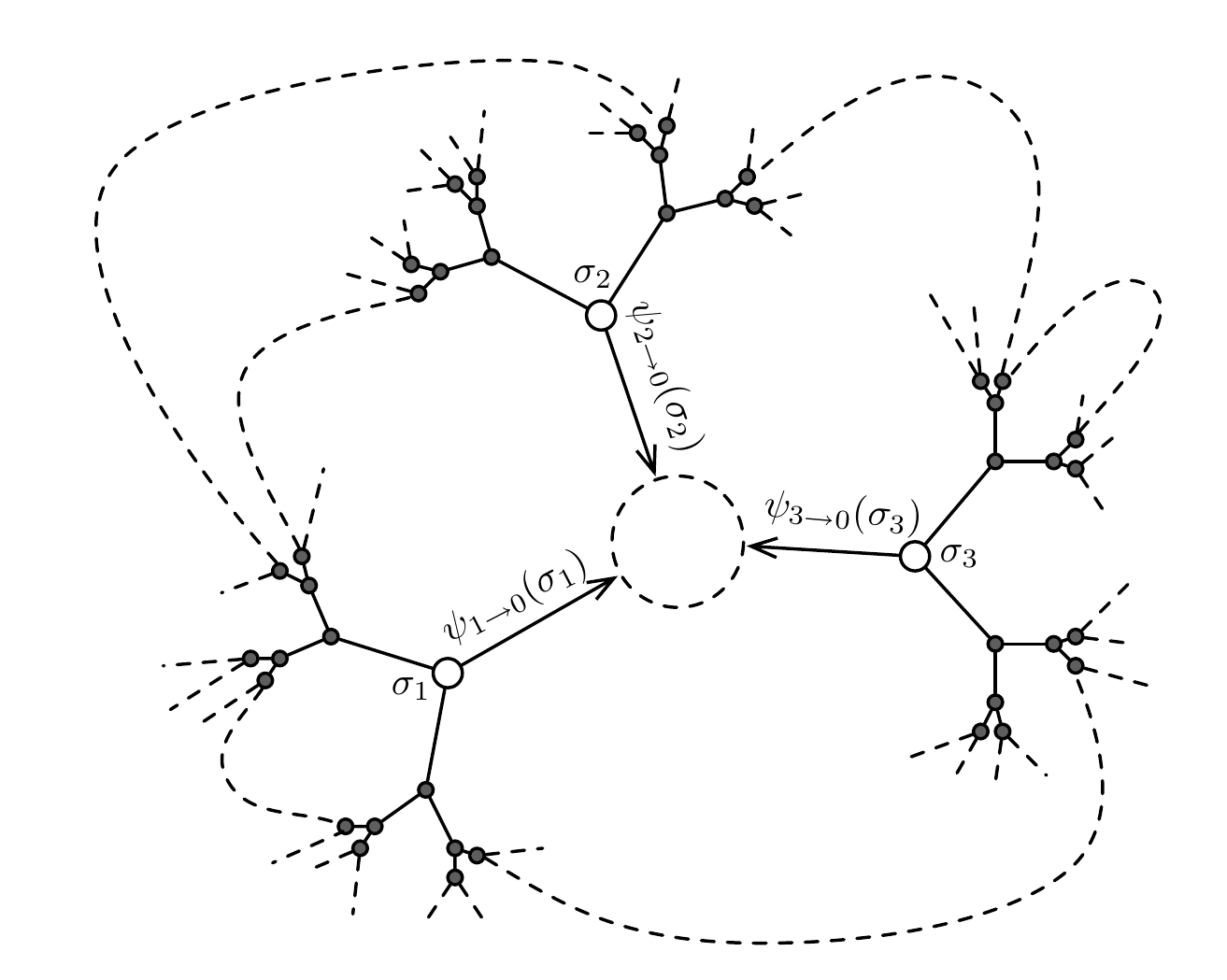}
		\includegraphics[width=2.5in]{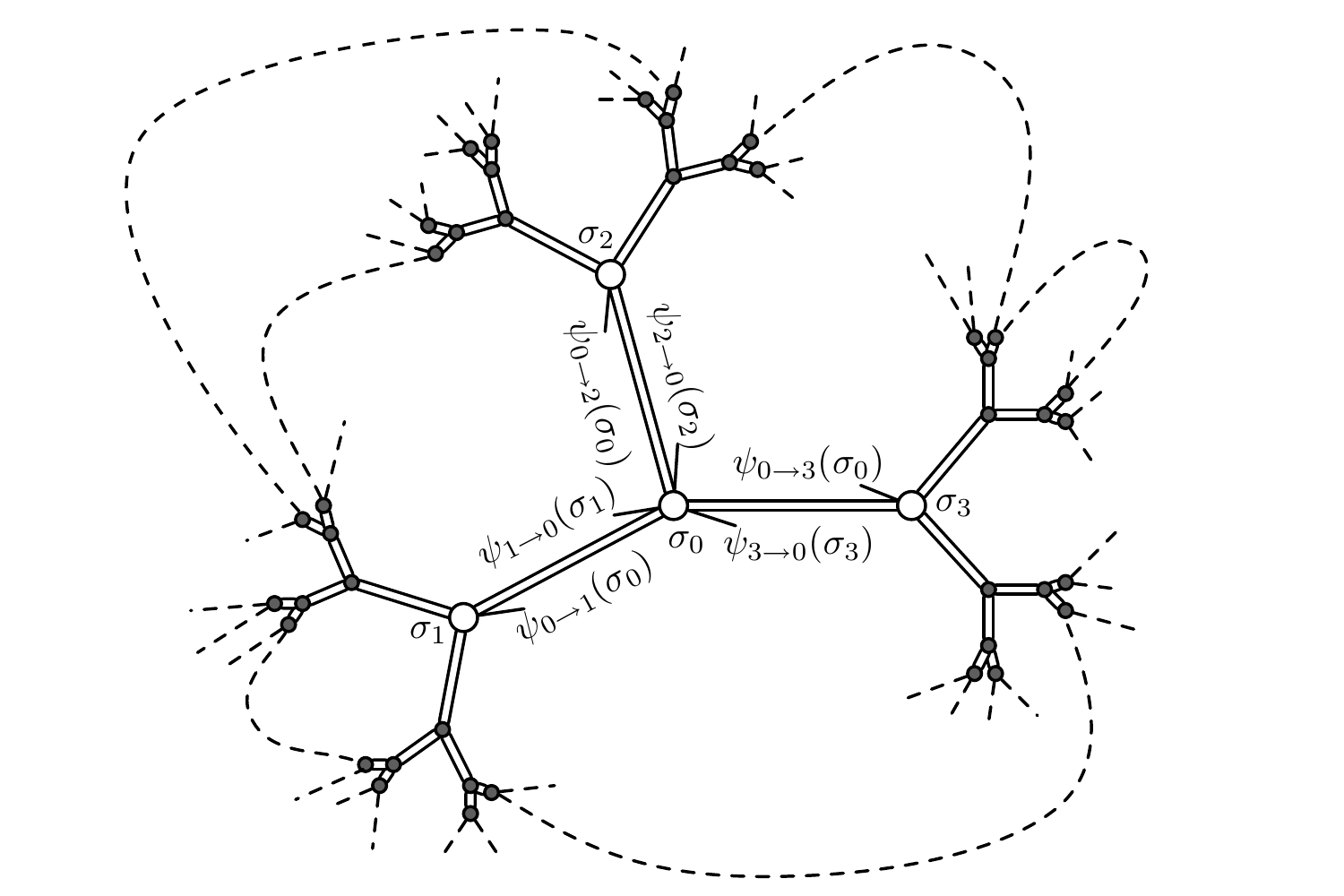}
	\caption{\emph{(Left)} Schematic of $q=3$ regular random graph
          with a cavity carved out at spin $\sigma_0$. \emph{(Right)}
          The belief propagation equations for a graph $G$ involve
          $2M$ cavity distributions, or \emph{beliefs}, $\psi_{i \to
            j}$, one for each of the two directions of a link in the
          graph.} 
	\label{fig:cavity-bp}
\end{figure}

Let us make these considerations more precise. Consider carving a cavity into
a large regular random graph $G$ with $N$ spins and $M$ edges representing two body interactions, as in
Fig.~\ref{fig:cavity-bp}. We will not explicitly consider the 
straightforward generalization to $k$-body interactions with $k>2$; it
needlessly complicates the notation. The statistical connection at
temperature $1/\beta$\footnote{We may eventually take the temperature
  to 0 ($\beta \to \infty$) in order to find actual ground state
  solutions of the optimization problem.} 
between the removed spin $\sigma_0$ and the rest of the graph is entirely
mediated by the \emph{joint} cavity distribution
\begin{equation}
	\psi_{G \backslash \set{0}}(\sigma_1, \sigma_2, \sigma_3) =
        \frac{1}{Z_{G\backslash\set{0}}} \prod_{j\notin \{0,1,2,3\}}
        \sum_{\sigma_j} e^{-\beta H_{G \backslash \set{0}}} 
\end{equation}
That is, the thermal distribution for $\sigma_0$ in the original model is given by:
\begin{equation}
	\psi_0(\sigma_0) = \frac{1}{Z_0} \sum_{\sigma_1,\sigma_2,
          \sigma_3} e^{-\beta (H_{01}+H_{02}+H_{03})} \psi_{G
          \backslash \set{0}} (\sigma_1, \sigma_2, \sigma_3) 
\end{equation}
If, on carving out the cavity, the neighboring spins become
independent then the joint cavity distribution factors: 
\begin{equation}
	\label{eq:cavity-ind}
	\psi_{G \backslash \set{0}}(\sigma_1, \sigma_2, \sigma_3) =
        \psi_{1\to 0}(\sigma_1) \psi_{2 \to 0}(\sigma_2) \psi_{3\to
          0}(\sigma_3) 
\end{equation}
On trees, the independence is exact because the cavity spins sit in disconnected clusters; on locally tree-like graphs, the cavity spins are only connected through long (divergent in system size) paths and thus we might expect Eq.~\eqref{eq:cavity-ind} to hold approximately. 

Thus the objects of interest are the $2M$ cavity distributions
$\psi_{i \to j}$, see Fig.~\ref{fig:cavity-bp}. These are also known
as messages or beliefs: $\psi_{i \to j}(\sigma_i)$ is a message passed
from site $i$ to site $j$ which indicates site $i$'s beliefs about
what it should do in the absence of site $j$, and thus also its
beliefs about what site $j$ should do to optimize the free
energy. Crucially, when the cavity distributions are independent, they
also satisfy the iteration relation: 
\begin{equation}
	\label{eq:bp-eqn}
	 \psi_{i \to j}(\sigma_i) = \frac{1}{Z_{i\to j}}
	 \prod_{k \in \partial i \backslash \set{j}} \sum_{\sigma_k} e^{-\beta H_{ik}(\sigma_i, \sigma_k)} \psi_{k \to j} (\sigma_k)
\end{equation}
These are the Belief Propagation (BP) equations and they are really
just the Bethe-Peierls self-consistency equations in more formal,
generalized, notation. Indeed, if we parameterize the functions
$\psi_{i\to j}$ by cavity fields $h_{i\to j}$ 
\begin{equation}
\psi_{i\to j}(\sigma_i)=\frac{1}{2\cosh(\beta h_{i \to j})}e^{-\beta h_{i \to j} \sigma_i},
\end{equation}
and specialize to an Ising Hamiltonian $H = \sum_{\langle ij \rangle}
J_{ij} \sigma_i \sigma_j$, then the BP equation becomes: 
\begin{equation}
	h_{i\to j} = \frac{1}{\beta}\sum_{k \in \partial i \backslash
          \set{j}} \tanh^{-1}\left[ \tanh(\beta J_{ki}) \tanh(\beta
          h_{k \to i})\right] 
\end{equation}
which should be familiar from Ising mean field theory.

There are linearly many BP equations, each of which is a simple relation
involving a finite summing out procedure on the right to define a cavity
distribution on the left. We may now consider two approaches to solving them:
a) take a thermodynamic limit and find the statistics of their solutions; and
b) iteratively solve them for finite $N$ using a computer. The former approach
leads to the so-called cavity equations and the estimates regarding the
thermodynamic phase diagram of Fig.~\ref{fig:figs_pnas-2007-ksat-pd}. 
The latter leads to the
belief propagation algorithm for solving particular instances of optimization
problems. Indeed, once the solution of the BP equations is known for a
particular instance, one can obtain a solution of the SAT formula by using a
decimation heuristic which fixes the variables with positive (resp.\ negative)
total of incoming messages to 1 (resp.\ 0) and re-running BP on the remaining
formula if necessary \cite{Braunstein:2005p110}.

There are as many BP equations as unknown cavity distributions and thus we
generically expect to find discrete solutions. However, there can be more than
one solution. For instance in the low temperature phase of the Ising
ferromagnet, there are three: the (unstable) paramagnet and the two (symmetry
related) magnetized solutions. For spin glass models with quenched disorder,
there may be exponentially many solutions, each corresponding to a
macroscopically distinct magnetization pattern in the system. When this
occurs, the belief propagation algorithm may fail to converge. In this case,
one needs to take into account the presence of multiple solutions of the BP
equations, which can be done statistically using an algorithm known as Survey
Propagation and in the thermodynamic limit using the `replica symmetry
breaking' cavity equations, which arise as a hierarchy of distributional
equations. These equations describe the statistics of solutions of the BP
equations. Although these analyses are important for a correct understanding
of many types of glassy optimization problems on tree-like
graphs~\cite{Mezard:2009vn}, we will not consider these technical
generalizations further. We note that the jargon of `replica symmetry
breaking' arises in a completely different approach to solving mean field
glasses based on the so-called replica trick and Parisi ansatz. The terms have
no intrinsic meaning in the context of the cavity approach.

For quantum stoquastic (or Frobenius-Perron) Hamiltonians -- those for
which a basis is known for which all off-diagonal matrix elements are
negative, which guarantees that there exists a ground state
wavefunction for which all components can be chosen to have a positive
amplitude -- BP has been generalized in some recent works
\cite{Laumann:2008zl,Hastings:2007p5560,Leifer:2008pg1899} and has
since been taken up in the study of a number of quantum models on
trees
\cite{Semerjian:2009p7636,Krzakala:2008dz,Carleo:2009p8162,Laumann:2010uq}. Spin-glasses
with a transverse field are a case in point. This might be relevant
for the study of the performance of some quantum adiabatic algorithms
for solving classical instances of $k$-SAT, which we turn to next. 


\subsection{Adiabatic quantum algorithm for classical $k$-SAT} 
\label{sub:adiabatic_quantum_algorithm_for_classical_k_sat}

Before we move on to the problem of quantum satisfiability
$k$-QSAT, let us first ask whether we could use a quantum algorithm to solve
classical $k$-SAT efficiently. One possible strategy for this is based on a
protocol known as adiabatic quantum computing \cite{Farhi:2001p802},
which in its full generality is equivalent to computation based on
circuits. Here we will discuss a particularly simple member of this
class of algorithms.  

Consider a time-dependent quantum Hamiltonian, with real time $t$ parametried
by $s (t)$ (with $0\le s \le 1$):
\begin{equation}
\HH (s) = (1 - s) \HH_\Gamma + s \HH_0~~~,
\label{Eq:7}
\end{equation}

\noindent where $\HH_\Gamma$ is a transverse-field term and
$\HH_0$ is obtained from Eq.~\ref{eq:3sat_energy} 
\begin{eqnarray}
\HH_\Gamma & = & -\Gamma \sum \sigma^{\rm x}_i \nonumber\\
\HH_0 & = & \sum_{m} E_{m} \left( \left\{
  \sigma^z_i \right\} \right)~~~, 
\label{Eq:8}
\end{eqnarray}

\noindent by replacing the $\sigma_i$'s by Pauli operators $\sigma^z_i$.

The ground state space of $\HH(s)$ at time $s = 1$ is
spanned by (one of) lowest-energy classical configurations (ground states) of the
$k$-SAT problem; it is these which we are after, but which can be hard to
find. By contrast, at time $s = 0$, the quantum ground state has all spins
polarized in the {\rm x}-direction. This is both easy to describe and to
prepare. If we start the system in its ground state at $s = 0$, and
change $\HH(s)$ sufficiently slowly, the state of the system will
evolve adiabatically, and reach the desired state at time $s = 1$. 

However, we do not want to change $\HH(s)$ arbitrarily slowly, as
this would be no gain over an exponentially long classical run-time.

What is it that limits the sweep-rate? A non-zero sweep rate can induce
transitions to excited states. Careful derivation of the adiabatic
theorem  reveals that the probability of nonadiabatic transitions
tends to zero so long as the sweep rate is slower than the minimal
adiabatic gap $\Delta$ squared\footnote{In fact, there is some
  controversy in the rigorous literature about whether the asymptotic
  sweep rate must be slower than $1/\Delta^2$ or $1/\Delta^{2+\delta}$
  for some arbitrarily small constant $\delta$. See
  \cite{Aharonov:2008p7450}.}. That is, the run time $T$ must be
greater than or of order $O(1/\Delta^2)$ in order to ensure
adiabaticity.  

Heuristically, we need to be concerned about avoided level crossings in the
course of the evolution and in particular, the avoided crossing at the location of the minimal gap $\Delta$.%
\footnote{In the absence of any symmetries and fine-tuning, all level 
  crossings are avoided as a function of a single adiabatic parameter
  $s$.} 
As two levels approach closely, we get an effective two-level problem:
\begin{equation}
\HH_2 = \left(
\begin{array}{c@{\quad \quad}c}
\alpha(t - t_0) & \Delta/2 \\
\Delta/2 & -\alpha(t - t_0) 
\end{array}
\right)~~~.
\label{Eq:9}
\end{equation}

\noindent At the closest approach, at $t = t_0$, the ground state is separated
from the excited state by a gap $\Delta$. For $|\alpha(t-t_0)| \gg \Delta$,
variation of $t$ has very little effect on the adiabatic eigenstates and the
Schr\"odinger evolution remains adiabatic even for fast sweeping. It is only
the time spent in the interaction region $|\alpha(t - t_0)| < \Delta$ where
the adiabatic states rotate significantly and nonadiabatic transitions may
arise. Thus, the interaction time is $t_I \sim \Delta / \alpha$ and the
dimensionless figure of merit for adiabatic behavior should be $\Delta \cdot
t_I \sim \Delta^2 / \alpha$. In particular, for low sweep rates $\alpha \ll
\Delta^2$ or long run times $T \ge O(1/\Delta^2)$, we expect to have purely
adiabatic evolution. We note that the nonadiabatic transition probability $P$
for this two-level model was calculated some eighty years ago by Landau
\cite{Landau:1932fk} and Zener\cite{Zener:1932uq} whose exact result:
\begin{equation}
	P = 1 - e^{-\pi \Delta^2 / 4 \hbar \alpha}
\end{equation} 
quantifies the physical intuition in this case.

The quantum adiabatic algorithm has been studied extensively since its
introduction ten years ago on a number of hard random constraint
satisfaction problems closely related to
3-SAT~\cite{Smelyanskiy:2004p9667,Farhi:2001p802}. The critical
question is simple: how does the typical minimal gap encountered
during the procedure scale with increasing instance size $N$? Analytic
work on simple (classically easy) problem ensembles found several
polynomial time quantum adiabatic algorithms. Moreover, early
numerical studies for very small instances of harder problems held
promise that the gap would scale only polynomially
\cite{Young:2008p8866,Farhi:2001p802}. Unfortunately, subsequent
numerical studies on larger systems indicate that the gap eventually
becomes exponentially small due to a first order transition between a
quantum paramagnet and a glassy state~\cite{Young:2010p8867}. Even
worse, recent perturbative work argues that many-body localization
leads to a plethora of exponentially narrow avoided crossings
throughout the glassy phase
\cite{Altshuler:2009p8898,Altshuler:2009p8728} and thus the algorithm
discussed here does not produce an efficient solution of random 3-SAT
or related random constraint satisfaction problems. We note that this
is consistent with other evidence that quantum computers will not
enable the efficient solution of NP-complete problems. 



\section{Statistical mechanics of random $k$-QSAT} 
\label{sec:statistical_mechanics_of_random_k_qsat}

Let us now turn to the study of random instances of quantum satisfiability
$k$-QSAT. As discussed in Section~\ref{sub:qma_completeness_kitaev},
$k$-QSAT is QMA$_1$-complete and thus should be generally intractable. As in
the classical case, one might hope to gain some insight into the nature of the
difficulty of the quantum problem by studying a random ensemble of its
instances. Moreover, the richness of phenomena exhibited by the classical
random satisfiability problem -- and the many important spin-off techniques
that have been developed in their study -- encourages us to seek analogous
behaviors hiding in the quantum system.

\subsection{Random $k$-QSAT ensemble} 
\label{sub:random_k_qsat_ensemble}

Let us recap the definition of $k$-QSAT from Section~\ref{sub:qma_completeness_kitaev}:
\begin{description}
	
	\item[\bf Input:] A quantum Hamiltonian $H = \sum_m \Pi_m$ composed of $M$ projectors,
each acting on at most $k$ qubits of an $N$ qubit Hilbert space. 

  \item[\bf Promise:] Either $H$ has a zero energy state or all states
    have energy above a promise gap $\Delta > 1 /\textrm{poly}(N)$. 

  \item[\bf Question:] Does $H$ have a zero energy ground state?

\end{description}
Quantum satisfiability is a natural generalization of the classical
satisfiability problem: bits become qubits and $k$-body clauses become $k$-body
projectors. In key contrast to the classical case, where the binary variables
and clauses take on discrete values, their quantum generalizations are
continuous: the states of a qubit live in Hilbert space, which allows for
linear combinations of $|0\rangle$ and $|1\rangle$. 

Thinking of a Boolean
clause as forbidding one out of $2^k$ configurations leads to its quantum
generalization as a projector $\Pi^I_\phi \equiv | \phi \rangle\langle \phi |$,
which penalizes any overlap of a state $|\psi\rangle$ of the $k$ qubits in set
$I$ with a state $|\phi\rangle$ in their $2^k$ dimensional Hilbert space.
Indeed, if we restrict the $\Pi_m$ to project onto computational basis states,
$k$-QSAT reduces back to $k$-SAT -- all energy terms can be written as discrete
$0$ or $1$ functions of the basis state labels and the promise gap is
automatically satisfied since all energies are integers.

As in the classical problem, we make two random choices in order to specify an instance:
\begin{description}
\item[(R1)] each $k$-tuple occurs in $H$ with probability $p = \alpha N
  \left/ \left( N \atop k \right) \right.$
\item[(R2)] each $k$-tuple occurring in $H$ is assigned a projector
  $\Pi_m = |\phi\rangle\langle\phi|$, uniformly chosen from the space
  of projectors of rank $r$. For these notes, we will mostly consider
  the case $r=1$, although higher rank ensembles can be studied
  \cite{Movassagh:2010fk}. 
\end{description}

\noindent The first rule is identical to that of the classical random ensemble
and thus the geometry of the interaction graphs (Figure~\ref{fig:figs_graphs}) is
the same -- locally tree-like with long loops for sufficiently high clause
density $\alpha$.

The second rule, however, rather dramatically changes the nature of our random
ensemble -- the measure on instances is now continuous rather than discrete.
This turns out to be a major simplification for much of the analysis:
\emph{Generic} choices of projectors reduce quantum satisfiability to a graph,
rather than Hamiltonian, property. This ``geometrization'' property allows us
to make strong statements about the quantum satisfiability of Hamiltonians
associated with both random and non-random graphs and even non-generic choices
of projector, both analytically and numerically. For the remainder of these
notes, we use the term \emph{generic} to refer to the continuous choice of
projectors and \emph{random} to refer to the choice of the graph. See
Section~\ref{sub:geometrization_theorem} below for a more detailed discussion
of geometrization.


\subsection{Phase diagram} 
\label{sub:phase_diagram}

The first step in understanding the random ensemble is to compute the
statistics of this decision problem as a function of $\alpha$. Specifically we
would like to know if there are phase transitions in the satisfying manifold
as $\alpha$ is varied: these include both the basic SAT-UNSAT transition as
well as any transitions reflecting changes in the structure of the satisfying
state manifold. Additionally, we would like to check that the statistics in
the large $N$ limit are dominated by instances that automatically satisfy the
promise gap.

\begin{figure}[tbp]
	\centering
		\includegraphics[width=3.8in]{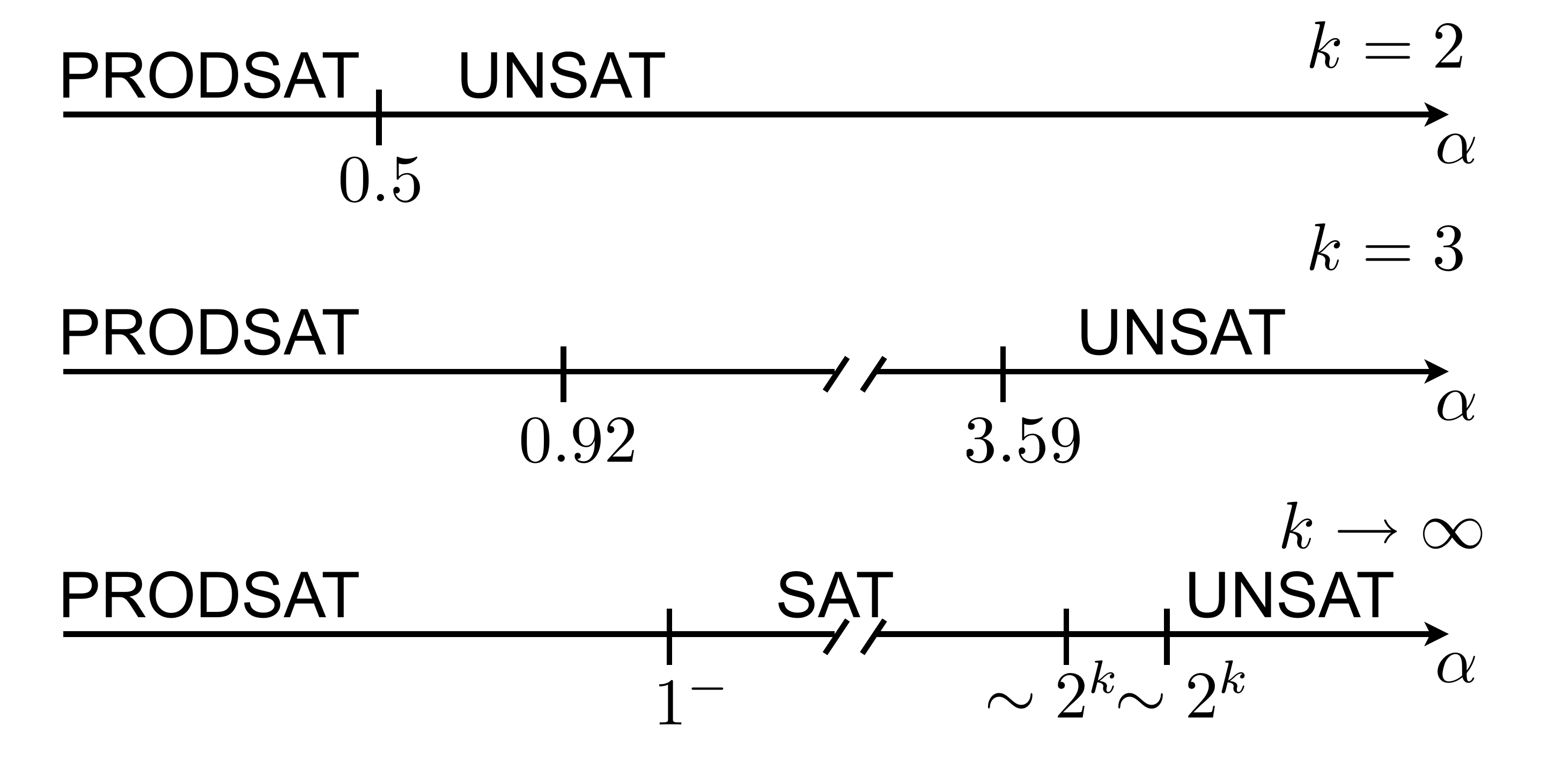}
	\caption{Phase diagram of $k$-QSAT.}
	\label{fig:qsat-pd}
\end{figure}

The current state-of-the-art QSAT phase diagram is shown in Figure
\ref{fig:qsat-pd}. Let us walk through a few of the features
indicated. First, we have separated the $k=2$ case from the higher
connectivity cases because it is significantly simpler. For $k=2$, we
can solve the satisfiability phase diagram rigorously and even
estimate energy exponents for the (non-zero) ground state energy above
the satisfiability transition. Our ability to do so is consistent with
the fact that 2-QSAT is in P -- instances of 2-QSAT can be efficiently
decided by \emph{classical} computers! In particular, it can be shown
that the zero energy subspace can be spanned, if it is nontrivial, by
\emph{product} states. Since these are much simpler to specify
classically (a product state needs only $2N$ complex numbers instead
of $2^N$), it is perhaps not surprising that we can decide whether or
not such states exist that satisfy a given instance $H$. In any event,
the only significant feature of the phase diagram is that for $\alpha
< \alpha_s = 1/2$, we have a PRODSAT phase -- that is a phase which is
satisfiable by unentangled product states -- and for $\alpha >
\alpha_s$, we have an UNSAT phase with finite ground state energy
density. The transition coincides with a geometric transition:
$\alpha_s = 1/2$ corresponds to the emergence of a giant component
in the underlying interaction graph. 

For $k \ge 3$, the phase diagram is somewhat more interesting. Again,
at low $\alpha < \alpha_{ps}(k) \sim 1$ there exists a PRODSAT regime
in which satisfying product states are guaranteed to exist. Above
$\alpha_{ps}$, there are no satisfying product states, but the system
remains SAT -- thus, there is an ``entanglement'' transition in the
ground state space as a function of $\alpha$. Finally, above some
$\alpha_{c} \sim 2^k$, there is an UNSAT phase in which it can be
shown that there are no zero energy satisfying states. We note that
the emergence of a giant component in the underlying interaction graph
happens at $\alpha_{gc} = \frac{1}{k(k-1)} \ll \alpha_{ps} \ll
\alpha_{c}$ -- the various relevant transitions are well-separated at
large $k$. A variety of different techniques go in to showing the
existence of these transitions and phases -- we sketch a few of
these arguments in
Section~\ref{sub:a_few_details_of_phases_and_transitions}. 


\subsection{Geometrization theorem} 
\label{sub:geometrization_theorem}

One of the most useful tools for studying random quantum satisfiability is geometrization. 
That is, the satisfiability of a generic instance of QSAT is a purely geometric property of the 
underlying interaction graph. This point of view extends to many properties of generic 
instances of QSAT -- such as whether they are product satisfiable or not. Results about 
generic QSAT thus follow from identifying the right geometric properties in the interaction 
graph ensemble. We discuss a few of these
examples in the sections to follow. Here, we provide an elementary
proof of geometrization and a few immediate corollaries. 

\newtheorem{geothm}{Geometrization Theorem}

\vspace*{12pt}
\begin{geothm}
\label{thm:geometrization}
  Given an instance $H$ of random k-QSAT with interaction graph $G$, the
  degeneracy of zero energy states $R(H) = \dim(\ker(H))$ takes a particular
  value $R_G$ for almost all choices of clause projectors. $R_G$ is minimal
  with respect to the choice of projectors.
\end{geothm}
\vspace*{12pt}

\proof{
  For a fixed interaction graph $G$ with $M$ clauses, $H = H_\phi =
  \sum_{i=1}^{M} \Pi_i = \sum_{i=1}^{M} \ket{\phi_i}\bra{\phi_i}$ is a
  matrix valued function of the $2^k M$ components of the set of $M$
  vectors $\ket{\phi_i}$. In particular, its entries are polynomials
  in those components. Choose $\ket{\phi}$ such that $H$ has maximal
  rank $D$. Then there exists an $D\times D$ submatrix of $H$ such
  that $\det(H|_{D\times D})$ is nonzero. But this submatrix
  determinant is a polynomial in the components of $\ket{\phi}$ and
  therefore is only zero on a submanifold of the $\ket{\phi}$ of
  codimension at least 1. Hence, generically $H$ has rank $D$
  and the degeneracy $R_G = \dim(\ker(H)) = 2^N - D$. \qed
}

The theorem holds for general rank $r$ problems as well by a simple
modification of the argument to allow extra $\phi$'s to be associated
to each edge. 

A nice corollary of this result is an upper bound on the size of the SAT phase
at any $k$. Consider any assignment of classical clauses on a given interaction
graph: this is a special instance of $k$-QSAT where the
projectors are all diagonal in the computational basis. As this is a
non-generic choice of projectors, the dimension of its satisfying manifold is
an upper bound on the dimension for generic choices. We conclude then that the
classical UNSAT threshold is an upper bound on the quantum threshold. Indeed,
if we can identify the most frustrated assignment of classical clauses on a
given interaction graph, i.e. the assignment that minimizes the number of satisfying
assignments, we could derive an even tighter bound.

\newtheorem{corollary}{Corollary}

\vspace*{12pt}
\begin{corollary}
\label{thm:class-bounds-quantum-nullity}
  The generic zero state degeneracy is bounded
  above by the number of satisfying assignments of the most
  constrained classical $k$-SAT problem on the same graph.
\end{corollary}
\vspace*{12pt}

It is easy to construct example instances in which the quantum problem has
fewer ground states than the most frustrated classical problem on the
same interaction graph. Thus, the bound of corollary
\ref{thm:class-bounds-quantum-nullity} is not tight. 


\subsection{A few details of phases and transitions} 
\label{sub:a_few_details_of_phases_and_transitions}

There are three flavors of arguments that have been used to pin down
the phase diagram of $k\ge 3$-QSAT: construction of satisfying product
states \cite{Laumann:2010kx,Laumann:2010p7275}; combinatorial upper
bounds on the zero state
degeneracy\cite{Bravyi:2009p7817,Laumann:2010p7275}; and, a
non-constructive invocation of the quantum Lovasz local lemma to
establish the entangled SAT phase\cite{Ambainis:2009fv}. All three
ultimately rely on establishing a correspondence between some
geometric feature of the interaction graph $G$ (or its subgraphs) and
the properties of zero energy states for generic instances through
geometrization. We sketch each of the three kinds of results below and
refer the motivated reader to the relevant literature for details. 

\subsubsection{Product states} 
\label{ssub:constructing_product_states}

Perhaps the most direct approach to establishing a SAT phase in the phase
diagram is to attempt explicitly to construct satisfying product
states  \cite{Laumann:2010p7275}. This is sufficient completely to 
determine the $k=2$ SAT phase, but only proves the existence of the
PRODSAT phase of the $k\ge3$ phase diagram. 

In this approach, one
constructs product states by a ``transfer matrix''-like procedure: if
there is a product state on some interaction graph $G$, then we can
extend it to a product state on a graph $G'$ which is $G$ plus one
additional clause $C$, so long as the clause $C$ has at least one
qubit not already in $G$. Thus, if a given graph $G$ can be built up
one clause at a time in an order such that there is always a
previously unconstrained qubit brought in by each additional clause,
$G$ is PRODSAT.  

Moving beyond our explicit construction, a complete characterization
of product satisfiability can be found by analyzing the equations
which a satisfying product state must obey \cite{Laumann:2010kx}. This
is a system of $M$ algebraic equations in $N$ complex unknowns and
naive constraint counting suggests that $M \le N$ should have
solutions while $M > N$ should not. Since the system is sparse,
however, a somewhat more detailed analysis is required to show: 

\newtheorem{theorem}{Theorem}
\begin{theorem}\label{thm:prodsat}
	$G$ is PRODSAT for generic choices of projectors $\Pi_m$ if
        and only if its interaction graph has a dimer covering of its
        clauses. 
\end{theorem}

Here a ``a dimer covering of its clauses'' is a pairing between qubits
and clauses such that every clause appears paired with exactly one
qubit and no qubit or clause appears more than once. The proof relies
on `product state perturbation theory', or in other words, the
smoothness of the complex manifolds defining the projector space and
the product state space. 

\begin{figure}[t]
	\centering
		\includegraphics[scale=1]{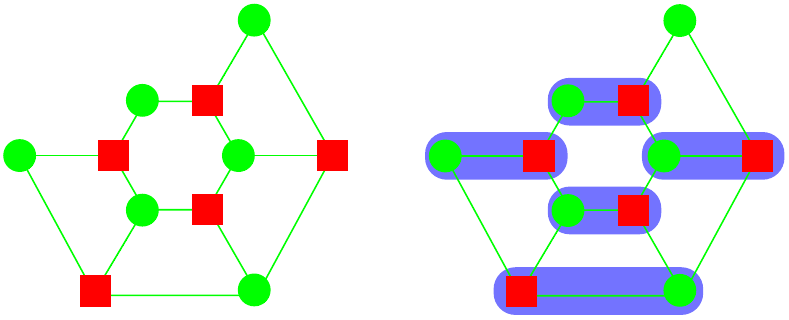}
	\caption{Example of a $k=3$ interaction graph with $M<N$,
          circles (green) indicate qubits and squares (red) indicate
          clause projectors that act on adjacent qubits (left); a
          dimer (blue shaded) covering that covers all clauses
          (right).} 
	\label{fig:dimercover}
\end{figure}

If we apply the dimer covering characterization to the random
interaction graph ensemble for $G$, we find the $\alpha_{ps}$
indicated in the phase diagram of Figure~\ref{fig:qsat-pd}. In
particular, for $\alpha < \alpha_{ps}$, such dimer coverings exist
w.p. 1 in the thermodynamic limit while for $\alpha > \alpha_{ps}$
they do not. The location of the geometric transition for the
existence of dimer coverings in known in the literature
\cite{Mezard:2003p5977}. Thus, for $\alpha > \alpha_{ps}$ there are no
satisfying product states, although there may still be satisfying
entangled states. 

We note that the dimer covering characterization of product states
provides an explicit mapping between dimer coverings and generic
product states. In the case $M=N$, this mapping is one-to-one and
provides a handle on counting the number of product states and, with
more work, the ability to estimate their linear dependence. There are
many avenues to explore here. 


\subsubsection{Bounding the degeneracy} 
\label{ssub:bounding_the_degeneracy}

The existence of an UNSAT regime for large $\alpha$ follows
immediately from the geometrization theorem and the existence of an
UNSAT regime for classical SAT. That is, for $\alpha > \alpha_s^{\rm
  Classical}$, typical graphs $G$ are classically UNSAT and therefore,
since the generic dimension $R_G$ of the zero energy state space is
minimal, they are also generically quantum UNSAT. In other words, the
quantum SAT-UNSAT transition $\alpha_s \le \alpha_s^{\rm Classical}$. 

This estimate of the SAT-UNSAT transition is not tight: the quantum
UNSAT phase begins at a lower $\alpha$ than the classical UNSAT
phase. This can be seen using another approach to bounding the ground
state degeneracy. In this approach, one builds up a given graph $G$
out of small clusters, each of which decimates the satisfying
eigenspace by some known fraction. Indeed, if we consider two
interaction graphs $K$ and $H$ on $N$ qubits with that respective generic 
zero energy dimensions $R_K$ and $R_H$, it is straightforward
to show \cite{Bravyi:2009p7817} 
\begin{equation}
	R_{K\cup H} \le R_K \frac{R_H}{2^{N}}
\end{equation}
for generic choices of projectors on $K$ and $H$. As an example, let
us build a graph $G$ with $M$ clauses, one clause at a time. Each
individual clause $H$ has $R_H = (1-1/2^k) 2^N$ because it penalizes 1
out of the $2^k$ states in the local $k$-qubit space and leaves the
other $2^{N-k}$ alone. Thus, adding each additional clause $H_m$
decimates the satisfying subspace by at least a factor $(1-1/2^k)$
and: 
\begin{equation}
	R_G \le 2^N (1-1/2^k)^M
\end{equation}
Plugging in $M = \alpha N$ and taking $N$ to infinity, we find that
for $\alpha > -1 / \log_2 (1-1/2^k)$, $R_G$ must go to zero. This
simply reproduces the Pauling bound mentioned for classical $k$-SAT in
Section~\ref{sub:classical_statistical_mechanics_of_k_sat}. 

However, one can do better by taking somewhat larger clusters $H$,
calculating $R_H$ exactly for these small clusters and then working
out how many such clusters appear in the random graph $G$. This leads
to much tighter bounds on $\alpha_s$ from above and in particular, as
shown in Ref.~\onlinecite{Bravyi:2009p7817}, $\alpha_s(k) <
\alpha_s^{\rm Classical}(k)$ for all connectivities $k$. 


\subsubsection{Quantum Lovasz local lemma} 
\label{ssub:quantum_lovasz_local_lemma}

The final technique that has been used to fill in the phase diagram of
Figure~\ref{fig:qsat-pd} is the development of a quantum version of the Lovasz
local lemma \cite{Ambainis:2009fv}. This lemma provides a nonconstructive
proof that satisfying states must exist for $k$-QSAT instances built out of
interaction graphs with sufficiently low connectivity -- that is, graphs in
which the degree of every qubit is bounded by $2^k/(e k)$. The QSAT ensemble
which we study in fact has average degree $\alpha k$ but the degree
distribution has an unbounded tail. By cutting the graph into low and high
connectivity subgraphs, using the product state characterization on the high
connectivity part and the Lovasz lemma on the low connectivity part, and
carefully glueing these results back together, it is then possible to show
that satisfying states exist for $\alpha < 2^k / (12 e k^2)$.

For sufficiently large $k$, this result proves that $\alpha_s \ge O(2^k / k^2)
\gg \alpha_{ps}$, establishing the entangled SAT regime indicated in
Figure~\ref{fig:qsat-pd}.

We now sketch the idea behind the classical and quantum Lovasz local lemmas. 

Suppose we have some classical probability space and a collection of $M$
events $B_m$, each with a probability of occurring $P(B_m) \le p < 1$. We
think of these as low probability `bad' events, such as ``the $m$'th clause of
a $k$-SAT instance is not satisfied by $\sigma$'' given a uniformly chosen
configuration $\sigma$. In this particular case, $P(B_m) = p = 1/2^k$ for all
clauses $m$. If there is a positive probability that no bad event comes to
pass, then there is clearly an overall assignment of $\sigma$ which satisfies
all of the clauses. Thus, we would like to show this probability is positive.

If the events $B_m$ are independent, this is clearly possible: 
\begin{equation}
	Pr(\bigwedge\limits_m \neg B_m) = \prod_{m=1}^M (1 - Pr(B_m)) \ge (1-p)^M > 0
\end{equation}
In the $k$-SAT example, clauses are independent if they do not share any bits
-- thus this argument provides us the rather obvious result that $k$-SAT
instances composed of only completely disconnected clauses are satisfiable. On
the other hand, if the events $B_m$ are dependent, it is clear that we can
make
\begin{equation}
	Pr(\bigwedge\limits_m \neg B_m) = 0.
\end{equation}
For instance, simply take a 3-SAT instance with 3 qubits and 8
clauses, each of which penalizes a different configuration. These
clauses still have individually low probability ($p = 1/2^k$) but at
least one of them is violated by any configuration. 

The classical Lovasz local lemma \cite{Erdos:1975vn,Moser:2009p8649}
provides an elementary method for relaxing the independence
requirement a little bit. In particular, if each event $B_m$ depends
on no more than $d$ other events $B_{m'}$ where 
\begin{equation}
	p\, e\, d \le 1
\end{equation}
(Euler's constant $e \approx 2.7182\dots$ being the basis of the natural logarithm) 
then the local lemma tells us that there is
indeed a positive probability that no bad event happens: 
\begin{equation}
	Pr(\bigwedge\limits_m \neg B_m) > 0
\end{equation}
This means that for connected $k$-SAT instances of sufficiently low
degree, Lovasz proves the existence of satisfying configurations. 

In the quantum generalization of the Lovasz lemma, probability is
replaced by the relative dimension of satisfying subspaces. That is,
for a QSAT projector $\Pi$ of rank 1, the ``probability of the clause
being satisfied'' is
\begin{equation}
\frac{\textrm{Dim}(SAT)}{\textrm{Dim}(\mathcal{H})}=\frac{2^k-1}{2^k}=
1 - \frac{1}{2^k}~~. 
\end{equation}
With the right definitions in hand, the
generalization is also elementary and the result looks nearly
identical to the classical case. However, now a positive probability
that all projectors are satisfied tells us that there exists a
(potentially quite entangled) quantum state of an $N$-qubit Hilbert
space which satisfies the instance of $k$-QSAT. 

The Lovasz local lemma is nonconstructive because it works by bounding (from
below) the satisfying space degeneracy as the graph $G$ is built up, so long
as each additional clause does not overlap too many other clauses. In some
sense this is dual to the arguments used to prove the UNSAT phase exists by
bounding this degeneracy from above, but the technical details are somewhat
more subtle since they require a more careful consideration of the interaction
between additional clauses and the existing constraints.

In the last few years, computer scientists have developed a
\emph{constructive} version of the classical Lovasz local lemma. That
is, there are now proofs that certain probabilistic algorithms will
actually efficiently construct the Lovasz satisfying states
\cite{Moser:2009p8649}. Recent work suggests that a quantum
generalization of this constructive approach may also be possible
\cite{Sattath:2010wip}. 



\subsection{Satisfying the promise} 
\label{sub:satisfying_the_promise}

Ideally, we would like to study an ensemble of $k$-QSAT instances which always
satisfy the promise. Such an ensemble would only contain Yes-instances with
strictly zero energy and No-instances with energy bounded away from zero
energy by a polynomially small promise gap. Such an ensemble is hard to
construct as one does not know \emph{a priori} which instances have zero
energy or not, let alone whether their energy might be exponentially small.
The best we can hope to do is choose a random ensemble in which the promise is
satisfied statistically -- perhaps with probability 1 in the thermodynamic
limit.

Physical arguments suggest that the $k$-QSAT ensemble that we study here
satisfies the promise in this statistical sense and for $k=2$ it can be
proven. On the SAT side of the phase diagram, all of the arguments that have
been constructed to date show the existence of strict zero-energy states in
the thermodynamic limit. These arguments all rely on geometrization: the
existence of generic zero energy states is a graph property and such
properties are either present or not in the thermodynamic limit of the random
graph at a given $\alpha$. Hence the zero energy phase as determined by such
arguments is a strictly zero energy phase.

As statistical physicists, we expect that the UNSAT phase of $k$-QSAT has
extensive ground state energy with relatively vanishing fluctuations for any
$k$. If this is true, the promise that $E\ge O(N^{-a})$ fails to be satisfied
only with exponentially small probability by Chebyshev's inequality. More
generally, so long as the average ground state energy is bounded below by a
polynomially small scale $E \ge O(N^{-b})$ with relatively vanishing
fluctuations, the promise will be violated with only exponentially small
probability for $a > b$.

For $k = 2$, it can shown rigorously that the expected ground state energy for
$\alpha > \alpha_s = 1/2$ is bounded below by a nearly extensive quantity
(\emph{i.e.} $E \ge O(N^{1-\epsilon})$ for any $\epsilon > 0$). Also, we know
that the SAT phase extends to $\alpha = 1/2$ because we can show that
satisfying zero energy product states exist up to this clause density. Thus,
the ensemble satisfies the promise with high probability in both phases. At
the critical point, things are not quite so clear, but one might expect
fluctuations around $E=0$ at the scale $O(\sqrt{N})$ so that if the promise
gap is chosen to be $O(N^{-1})$, the weight of the ensemble below the gap
scale goes to zero.


\subsection{Open questions} 
\label{sub:open_questions}

In closing let us take stock of where we are at in the analysis of QSAT with
the set of results on SAT as our template. First, the phase diagram clearly
needs more work starting with more precise estimates for the SAT-UNSAT
boundary. Within the SAT phase we have identified one phase transition where
the satisfying states go from being products to being entangled and the key
question is whether there are any others and whether they involve a clustering
of quantum states in some meaningful fashion. Second, we have not said
anything about the performance of algorithms for QSAT or about the
relationship between phase structure and algorithm performance. Apart from
some preliminary work on the adiabatic algorithm for 2-SAT
\cite{Govenius:2008fk}, this direction is wide open for exploration.



\section{Conclusion} 
\label{sec:conclusion}

In this review, we have tried to provide a reasonably self-contained introduction 
to the statistical mechanics of classical and quantum computational 
complexity, starting at the  venerable subject of classical complexity theory, 
and ending at an active current research frontier at the intersection in quantum 
computing, quantum complexity theory and quantum statistical mechanics. We 
hope that this review will not only encourage some of its readers to contribute 
to these fields of study, but that it will also have provided than with some of the 
background necessary for getting started.

\subsubsection*{Acknowledgements}

We very gratefully acknowledge collaborations with Andreas L\"auchli, in particular 
on the work reported in Ref. \cite{Laumann:2010kx}. 
Chris Laumann was partially supported by a travel award of ICAM-I2CAM under 
NSF grant DMR-0844115. 

\bibliographystyle{habbrv}
\bibliography{qsat}

\end{document}